\documentclass[journal,onecolumn, draftclsnofoot]{IEEEtran}
\usepackage{dsfont}
\usepackage{graphicx}
\usepackage{subfigure}
\usepackage{amsfonts,amssymb}
\usepackage{bm}
\usepackage{lipsum}
\usepackage{cite}
\usepackage{bm}
\usepackage{subeqnarray}
\usepackage{caption}
\usepackage{lipsum}
\usepackage{multicol}

\usepackage{amsfonts,amssymb}
\usepackage{lipsum}
\usepackage{bm}
\usepackage{subeqnarray}
\usepackage{color}
\usepackage{amsthm}
\usepackage{graphicx}
\usepackage[ruled]{algorithm2e}

\usepackage{array}
\newcommand{\PreserveBackslash}[1]{\let\temp=\\#1\let\\=\temp}
\newcolumntype{C}[1]{>{\PreserveBackslash\centering}p{#1}}
\newcolumntype{R}[1]{>{\PreserveBackslash\raggedleft}p{#1}}
\newcolumntype{L}[1]{>{\PreserveBackslash\raggedright}p{#1}}

\graphicspath{{figures/}}

\theoremstyle{plain}
\newtheorem{thm}{Theorem}
\newtheorem{lem}{Lemma}

\theoremstyle{definition}
\newtheorem{defn}{Definition}

\theoremstyle{remark}

\usepackage[cmex10]{amsmath}

\usepackage{amsfonts,amssymb}
\usepackage{epstopdf}
\usepackage{amsmath}
\usepackage{bm}
\usepackage{algorithmic}

\begin{document}
%


\title{Private and Truthful Aggregative Game for Large-Scale  Spectrum Sharing}

\author{\normalsize Pan Zhou, \emph{Member, IEEE}, Wenqi Wei, \emph{Student Member, IEEE}, Kaigui Bian, \emph{Member, IEEE}, Dapeng Oliver Wu, \emph{Fellow, IEEE}, Yuchong Hu, \emph{Member, IEEE}, Qian Wang, \emph{Member, IEEE}  
\thanks{Pan Zhou and Wenqi Wei, these authors contributed equally to this work and are considered co-first authors, are with School
of Electronic Information and Communications, Huazhong University of
Science and Technology, Wuhan 430074, China (email: panzhou@hust.edu.cn, wenqiweihust@hust.edu.cn).

Kaigui Bian is with the Institute
of Network Computing and Information Systems,
School of EECS,  Peking University, Beijing 100871, China (email: kaiguibian@gmail.com).

Dapeng Oliver Wu is with Department of Electrical
and Computer Engineering, University of Florida, Gainesville, FL
32611, USA (email: wu@ece.ufl.edu).

Yuchong Hu, School of Computer Science and Technology, Huazhong University of
Science and Technology, Wuhan 430074, China (email: yuchonghu@hust.edu.cn).


Qian Wang is with the Key Lab of Aerospace Information Security and Trusted Computing, School of Computer, Wuhan University, Wuhan 430072, China (email: qianwang@whu.edu.cn).

This work was supported by the National Science
Foundation of China under Grant 61401169, Grant CNS-1116970, and Grant
NSFC 61529101.
} } 

\maketitle
\thispagestyle{empty}
\pagestyle{plain}

\begin{abstract}
 Thanks to the rapid development of information technology, the size of wireless network is becoming larger and larger, which makes spectrum resources more precious than ever before. To improve the efficiency
  of spectrum utilization, game theory has been applied to study efficient spectrum sharing for a long time. However, the scale of wireless network in existing studies is relatively small. In this paper, we introduce a novel game called aggregative game and model spectrum sharing in a large-scale, heterogeneous, and dynamic network using such game concept. Meanwhile, the massive usage of spectrum leads to easier divulgence of privacy of spectrum users, which calls for privacy and truthfulness guarantees.
  In a large decentralized scenario, each user has no priori about other users' channel access decisions, which forms an incomplete information game. A ``weak mediator'', e.g., the base station or licensed spectrum regulator, is introduced and turns this game into a complete one, which is essential to reach a Nash equilibrium (NE). By utilizing past experience on the channel access, we propose an online learning algorithm to improve the utility of each user. We show that the learning algorithm achieves NE over time and provides  no regret guarantee for each user. Specifically, our mechanism admits an approximate \emph{ex-post}. The mechanism is also joint differentially private and is incentive-compatible. Efficiency of the approximate NE is evaluated, and innovative scaling law results  are
  disclosed. We also provide simulation results to verify  our analysis. 
\end{abstract}

\IEEEpeerreviewmaketitle


\section{Introduction}
\subsection{Motivation}
Dynamic spectrum sharing has been considered as a promising technique that allows unlicensed secondary users (SUs) to opportunistically access idle channels owned by legacy spectrum holders. In general, the increasing spectrum demand and frequent spectrum usage often lead  to large and uncertain network dynamics. Competition among users makes spectrum utilization inefficient due to potential interferences and massive data  packet collisions.

To improve individual efficiency, each user may speculatively conduct a mixed strategy of channel access over multiple channels. However, such a selfish behavior may result in more severe channel contentions and degrade the performance of the entire network. Given that the size of the network keeps growing nowadays, it gets more and more challenging to resolve this issue.

Classic spectrum sharing problem is usually studied for small scale networks. In this aspect, numerous efforts have been made to design  efficient spectrum utilization mechanisms based on game theory \cite{MY}\cite{CH}. However, large-scale spectrum sharing game with the heterogeneous individual impact is less understood, where different users have different interference impacts on the aggregative contention probability. The different actions of users, i.e., the mixed channel access probabilities,   may contribute to such heterogeneous impacts.  Moreover, it is implausible to study spectrum sharing in a large-scale wireless network with a priori, or with complete information. Users' decisions of channels are so decentralized that any user is incapable of collecting all of them, which leads to an incomplete information setting. The need to improve the efficiency of such heterogeneous large-scale spectrum sharing motivates us to explore a novel equilibrium solution.


\subsection{Aggregative Game with Weak Mediator}
To study the impact on the contention from heterogeneous individuals in a large-scale network, it requires a completely different view and some new equilibrium concepts.  Therefore, we model spectrum sharing in wireless network as an aggregative game.  Informally, aggregative game means that the payoff of each user is a function of the user's own action and the aggregator of all users's actions \cite{JENSEN}. With such a game model, we are able to identify users' heterogeneous actions and learn their effects on the contention. Since there are multiple channels, we arrive at a multi-dimensional aggregative game. To the best of our knowledge, this is the first work to study large-scale spectrum sharing under the aggregative game.

\begin{figure}
\centering
\includegraphics[scale=.75]{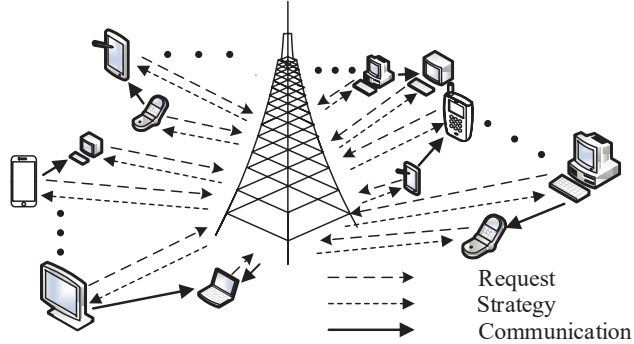}
\caption{Mediated large-scale multi-channel wireless network}
\label{fig:digraph1}
\vspace{-.2cm}
\end{figure}

Similar to \cite{NZD}, we model a network in which a couple of users share multiple channels. In the network,  users try to occupy channels with maximum effort to meet their spectrum demands. Such a competition makes spectrum sharing in large-scale wireless network a typical non-cooperative game \cite{NH1}.
   To form a complete information game and hence achieve a NE where every user can get his desired utility, we introduce a ``weak mediator" as it did in  \cite{RR}. The weaker mediator can be a base station (BS), a spectrum service anchor point or a licensed spectrum regulator.  Although users still lack the knowledge of others' choices of channels, the mediator can collect these channel access information and process them. As shown in Fig. \ref{fig:digraph1}, the BS works as the weaker mediator. We also assume that the mediator only has the power to collect users' reported mixed multi-channel access strategies, to calculate a mixed strategy profile and to provide non-binding channel access suggestions for each user. Such limited power enables us to call the mediator ``weak''.

Owing to the weak mediator, users can either opt-in using the BS as their advisor or opt-out neglecting its existence. For those users who are willing to opt-in, BS requires them to report their mixed channel access strategies. As for those who opt-out, e.g., devices which do not send their channel
 access ``requests" to BS in Fig. \ref{fig:digraph1}, they access the channel as they have intended. When BS gathers reported mixed strategies, it then calculates a mixed channel access strategy profile and suggests it to all users according to their submitted mixed strategies. Noted that even opting-out users receive the suggestions, but their suggestions are fixed because the action of opting-out is considered as a constant input. At last, all users access channels referring to or disregard the suggestion. Due to the weak mediator, even those who opt-in are able to neglect the suggestion. As a practical illustration,  the opt-in users could be high power wireless multimedia  users who are willing to report with little energy cost while the opt-out users could be
low power energy-constrained sensor nodes.

\subsection{Truthfulness and Privacy}
The mediated large-scale spectrum sharing forms a complete information game, but it raises two issues. On the one hand, the introduction of weak mediator calls for the \emph{truthfulness} guarantee. When BS cannot force any user to opt-in
or strictly follow its suggestions, users have the tendency to opt-out or opt-in telling lies. Two reasons account for such cheating. One reason is the incentive to get a higher payoff. The other reason is the fear of revealing their private information.


%
On the other hand, the \emph{privacy} issue in spectrum sharing arises when selfish users are competing \cite{HUANGQ}. When users are frequently accessing channels, they have a high risk of exposing their \emph{types}. A user's type, which can be inferred from his action, is his private
information. Such type can be the location of spectrum user, the purpose of spectrum usage, e.g. voice services, or watching a video, etc. The action is determined by his transmission power, contention probability and his mixed strategy over multiple channels. We will explain these
concepts in details later. Since a user's action is associated with his mixed strategy over multiple channels, it can be easily learned using historical statistics.
Moreover, in a busy wireless network, a high demand of spectrum require that users cannot hold channels for long. A user may only keep one channel for a certain period and need to compete for next usage. This makes inference about one's \emph{type} even easier.

Previously, cryptography was the main tool for preserving privacy in wireless networks \cite{Naor99}. But it often incurs high computation complexity. The recently proposed differential privacy \cite{DR} offers us an illuminating perspective. Differential privacy, intuitively, means that a single variation in the input data set can only have a limited impact on the output. The fact that differential privacy has superior performance in the large-scale data sets (i.e., user requests data sets) enlights an application in large-scale spectrum sharing. Besides, differential privacy is often implemented in a truthful manner \cite{ZLWS}. Thus, it is practical to achieve both truthfulness and privacy preservation via a differentially private mechanism. In this paper, we apply joint differential privacy \cite{KPRU}, which is adapted from standard
differential privacy,  to guarantee the privacy of spectrum users.
%
%
\subsection{Necessity of Online Learning}

The efficiency of NE based on one or a few sets of users' mixed channel access strategies varies. It sometimes suffers from users' speculations. In a large-scale spectrum sharing game, users need to repeatedly decide which channel to use. Therefore, we can
 utilize past accumulated information of channel choices of users to improve the efficiency of NE. We propose an online learning algorithm based on \cite{ARORA} to compute a mixed channel access strategy suggestion. The algorithm learns by maintaining weights on former mixed strategy experience. Through delicate design, we achieve an \emph{ex-post} NE  over time. The algorithm also gives no-regret guarantee to each user.

In fact, if users not only opt-in using the BS as the mediator but also truthfully report their types and subsequently follow the mediated instruction, the suggested strategy profile forms an approximate \emph{ex-post} NE of the mediated game. The resulting game achieves an approximate NE of the original complete information game. Although we can only achieve an approximate NE, we solve the equilibrium in strong sense if the spectrum sharing is in large scale. The approximation loss becomes less with the growing number of spectrum users,  which is promising in future large-scale networks.


\subsection{Main Contributions and Organizations}
In this paper, we design an efficient, truthful and privacy-preserving mechanism for large-scale spectrum sharing using aggregative game. The heterogeneous individual impact is also considered in the aggregative game model.
The main contributions of this work are described below.
\begin{itemize}
\item General game formulation: We formulate large-scale spectrum sharing in wireless network as an \emph{aggregative} game, with the heterogeneous individual impact on users' contention probability.

\item Truthfulness and privacy guarantee: The proposed mechanism preserves ($\varepsilon$,$\delta$)-joint differential privacy for each user and such privacy is implemented in a truthful manner. Algorithm $1$ prevents cheating learning in the BS and adversary users cannot distinguish a user's action by the NE he achieves. Besides, algorithm $3$ provides a privacy-preserving mixed strategy solution, which makes malicious statistics learning of actions through difficult.

\item Online learning for achieving NE: We propose an  online learning algorithm.
 The output of the algorithm, which is the mixed strategy suggestion profile, leads to an \emph{ex-post} NE over time.

\item Large-scale performance: The \emph{ex-post} NE achieved by online learning has better performance with the growing number of users, which demonstrates
promising scaling law results in the large-scale spectrum sharing.

\end{itemize}

The rest of this paper is organized as follows. We discuss related work in Section II. Section III gives aggregative game model of  multi-channel wireless network. Section IV presents several techniques of independent interest. Section V details the privacy-preserving and truthful design of our mediated spectrum
sharing game. Section VI gives online learning algorithm to compute NE. Numerical results are available in Section VII and the paper
is concluded in Section VIII.

\section{Related Work}

Numerous efforts have been made to design an efficient spectrum utilization mechanism based on game theory \cite{HUANG,NH2,Huang06,NH,MY,CH}. \cite{HUANG} showed that a pure strategy
equilibrium exists both for spatial spectrum access games on directed acyclic graphs and games satisfying the congestion property on directed trees and directed forests. \cite{NH} studied effective wireless network selection via evolutionary game approach while the author of \cite{MY} induced congestion game\cite{ROSE} into spectrum sharing. The classic congestion game assumes that a user's payoff depends on the number of users who share the same channels. The impact of individual users, which is related to the number of users competing for the same channel, is heterogeneous with regard to different users. \cite{CH} studied spatial congestion game and considered the heterogeneous individual impact. However, large-scale spectrum game with the heterogeneous individual impact taken into account is less understood. We here implement aggregative game \cite{JENSEN} to study heterogeneous spectrum sharing game.
To the best of our knowledge, this is the first to study large-scale heterogeneous spectrum sharing game under aggregative game.

Truthfulness has long been studied in terms of incentive compatibility in mechanism design \cite{AGT}, \cite{AMD}. We use both terms without difference. In spectrum sharing game, \cite{CLW} studied truthfulness in peer-to-peer network under spatial evolutionary game theory while \cite{FCZ} studied the truthful double auction mechanism for heterogeneous spectrums. The author of \cite{EPT} studied truthfulness in spectrum sharing problem where multiple systems coexist in an unlicensed band and interfere with each other.



As for privacy, McSherry and Talwar \cite{SMKT} first incorporated the techniques of differential privacy into mechanism design. Recently, Differential privacy has already been used in privacy-preserving wireless communications. Zhu and Shin \cite{ZS} first incorporated the techniques of exponential scheme \cite{SMKT} of differential privacy into mechanism design and studied cognitive radio spectrum auction. They implemented differential privacy in an auction in a truthful manner. In our large-scale wireless network, we use a number of tools developed from differential privacy \cite{DR}\cite{DMNA} to design a privacy-preserving spectrum sharing mechanism. We also adapt the online learning from \cite{ZLWS} to ensure that the mixed strategy output of every user is joint differential private \cite{KPRU}.
\section{Network Model}


This section presents an aggregative game model for the mediated spectrum sharing in wireless networks. Consider a wireless network where $n$ aggressive users are competing for $k$ channels. The user set is $\mathcal{N}=\{1, ...,i,..., n\}$ and the channel set is $\mathcal{K}=\{1, ...,d,..., k\}$. We also give the action set $\mathcal{M}=\{1, ...,j,..., m\}$, where a user's action is an application related index that specifies  its concrete  transmission power $\phi _{ij}^{d}$ and its contention probability $p_{i}$. Since transmission power varies when terminal users are performing different tasks on their devices, we associate the concept with the term action. e.g, transmission power needed for watching videos on the phone is different from that of making a phone call. Meanwhile, contention probability can reflect what the user is doing as well. A formal constitution of action is provided later.
According to Shannon capacity, each user $i$ with action $j$ over channel $d$ has a data rate $C_{ij}^d$:
\begin{equation}
\vspace{-.1cm}
C_{ij}^d = B_{i}^{d} \log _2 (1 + \frac{{\phi _{ij}^{d} \emph{g}^d_{i} }}{{ \omega _{i}^d }}),
\end{equation}
where $B_{i}^{d}$ is the spectrum bandwidth that user $i$ can get from channel $d$.  $\phi _{ij}^{d}$ is the fixed transmission power of user $i$'s action $j$ on channel $d$. $\omega^d _{i}$ denotes the background noise power while $\emph{g}^d_{i}$ is the channel gain of user $i$.


W.l.o.g.,  we define following individual
throughput as the utility function of user $i$ on channel $d$ under action $j$:
\begin{equation}
\vspace{-.1cm}
u_{ij}^d(p_i, \vec p)= C_{ij}^dp_i \prod\nolimits_{l \ne i} {(1 - p_{l} )},
\end{equation}
where $\vec p=\{ p_1, ..., p_i,..., p_n\}$ is the channel contention probability vector for
 all users. Taking fairness in communication into consideration, we define and write individual utility function $U_{ij}^d(p_i,\vec p)$ in a proportional-fair way:
\begin{equation}
U_{ij}^d(p_i,\vec p) = \log u_{ij}^d(p_i, \vec p) = \log C_{ij}^dp_{i} \prod\nolimits_{l \ne i} {(1 - p_{l} )}=\log C_{ij}^d + \log p_{i} + \sum\nolimits_{l \ne i}\log{(1 - p_{l} )}.
\label{aggre1}
\end{equation}

From equation (\ref{aggre1}), it is observed that user $i$'s utility function on channel $d$ is related to $\log p_{i} + \sum\nolimits_{l \ne i}\log{(1 - p_{l} )}$, which is an aggregative function of every competing user's contention probability on the channel. Noted that the contention probability of users who are not competing for the channel is ``0". Then we can define following aggregative function:
\begin{equation}
\vspace{-.2cm}
Q_{i}^d(\vec p) = \sum\nolimits_{l = 1}^n {q_{i}^d(p_l)},
\end{equation}
where $q_{i}^d(p_l)$ is the individual function for contention probability from user $i$'s perspective.
We here normalize  these $q_{i}^d(p_l)$ to be within $[0,1]$, and write it as:
\begin{equation}
q_{i}^d(p_l) = \left\{ \begin{array}{l}
 \frac{{\log p_{i}}}{{\log p_{i} + \sum\nolimits_{l \ne i} {\log (1 - p_l )} }},\quad l=i,\\
 \frac{{\log (1 - p_{l} )}}{{\log p_{i} + \sum\nolimits_{l \ne i} {\log (1 - p_l )} }},\quad l \ne i, \quad \forall l\in \mathcal{N}.\\
 \end{array} \right.
\end{equation}

It is explicit that each user has a heterogeneous impact on the aggregative contention probability function.
Thus, the utility function can be  written as  $U_{ij}^d(p_i,Q_{i}^d(\vec p)).$

\begin{figure}
\centering
\includegraphics[scale=.60]{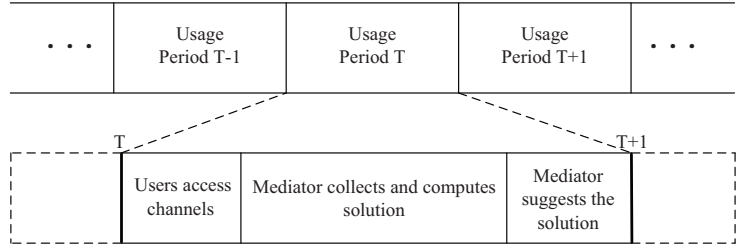}
\caption{Mediated process within one period}
\label{fig:digraph}
\end{figure}

Due to the limitation of power, the contention that a single user can cause to other users  should be limited for scalable network planning in large-scale spectrum sharing.  Accordingly, we introduce $\gamma$-aggregative game \cite{Babi}. This means that the greatest change a user can unilaterally cause to the aggregative contention probability is bounded by $\gamma$, which is determined as a scaling law index in later sections.
We can then revise the aggregative contention probability function by
\begin{equation}
\vspace{-.2cm}
Q_{i}^d(\vec p) = \gamma \sum\nolimits_{l = 1}^n {q_{i}^d(p_l)}.
\end{equation}
With the normalization and the assumption of $\gamma$-aggregative game,  the range of each channel's aggregative contention probability function $Q_{i}^d(\vec p) $ is bounded by $Q_{i}^d(\vec p) \in [0,n\gamma]$.


 According to formal  game theoretical  notation, we define $Q_{i}^d(p_i',\vec p_{-i})$, where $p_i'$ means that user $i$ deviates from his original action  given the
unchanged contention probabilities of other users, i.e., we have $\vec p_{-i}= \vec p-p_i$. Then, the following property is immediate:
 $$|U_{ij}^d(p_i, Q_{i}^d(\vec p)) - U_{ij}^d(p_i', Q_{i}^d(p_i',\vec p_{-i}))| = |Q_{i}^d(\vec p) - Q_{i}^d(p_i', \vec p_{-i}) )|.\vspace{-.1cm}$$
With the assumption of $\gamma$ aggregative game, utility functions are  $1$-Lipschitz with regard to the aggregative contention probability function,
$$\! |U_{ij}^d(p_i,\! Q_{i}^d(\vec p))\! - \! U_{ij}^d(p_i,\! Q_{i}^d(p_i',\vec p_{-i}))| \!\le\! ||Q_{i}^d(\vec p) \! - \!Q_{i}^d(p_i',\vec p_{-i}) )||_\infty.\vspace{-.1cm}$$

\begin{table}[tbp]
 \centering  
 \caption{Notations}
 \begin{tabular}{|C{1.3cm}|L{3.2cm}|C{1.9cm}|L{4.7cm}|}  
 \hline
 $\mathcal{N}$ & user set  & $\vec p$ & contention probability set\\ \hline  
 $\mathcal{M}$ & action set  & $\phi _{ij}^{d}$ & fixed transmission power  \\    \hline    
 $\mathcal{K}$ & channel set  & $\beta$ & confidence parameter \\ \hline
$\emph{g}^d_{i}$ & channel gain & $\omega^d _{i}$ & background noise power \\ \hline
 $B_{i}^{d}$ & spectrum bandwidth & $\alpha$ & discrete gap of aggregative functions\\ \hline
$C_{ij}^d$ & data rate & $\zeta$ & Lipschitz game approximation \\ \hline
 $\gamma$ & aggregative bound& $T$ & cost threshold in Alg.1\\ \hline
 $\vec P_d$ & mixed strategy set & $\varepsilon_0$, $\varepsilon$, $\delta$ & privacy parameters\\ \hline

  \end{tabular}
  \begin{tabular}{|C{2.3cm}|L{9.65cm}|}

 $\{P_{ij}^1,...,P_{ij}^k\}$ & mixed strategy set of user $i$ with action $j$ \\ \hline
$\pi$ & specific action defined by $\{\phi _{ij}^{d}$, $p_i$, $P_{ij}^{d}\}$ \\ \hline
 $q_{i}^d(p_l)$ & the individual function for contention probability from user $i$'s perspective \\ \hline
 $Q_{ij}^d( \vec p,\vec P_d)$ & aggregative contention probability function of user $i$ whose action $j$ on channel $d$ \\ \hline

 $ U_{ij}^d(p_i,Q_{i}^d(\vec p, \vec P_d)) $ & utility function for user $i$ with action $j$ on channel $d$ \\ \hline
$U_{ij} (p_i ,Q_i^d (\vec p,\vec P))$ & expected utility for user $i$ \\ \hline
 $\xi \!\!- \!\!\vec {BA}_{i}(Q_{i}^d(\vec p, \vec P_d)) $ & $\xi$-aggregative best response \\ \hline
 $\widehat Q_{ij}^d( \vec p,\vec P_d)$ & optimal set-valued aggregative function \\ \hline
 $\xi$ & approximation of aggregative best response \\ \hline
 $a(\widehat Q_{ij}^d( \vec p,\vec P_d))$ & difference between user $i$'s aggregative functions and set-valued function \\ \hline
 $E1,E2$ & approximation brought by Alg. 1 and Alg. 3 \\ \hline
 $\vec {P_i^t}$ & user $i$'s mixed channel access strategy $(P_{ij}^1,...,P_{ij}^k) $ at period $t$ \\ \hline
 ${\vec P}^t$ & mixed strategy profile $({\vec P}_1^t  ,...,{\vec P}_n^t  )$ of all users at period $t$
\\ \hline

 \end{tabular}
 \end{table}

When consider mixed strategy over multiple channels, we note that a user's  action is also determined by his
 mixed strategy. Specifically, a user who performs action $j$ conducts mixed strategy over all channels $\{P_{ij}^1,...,P_{ij}^k\}$ each with a probability $P_{ij}^d, \forall i \in \mathcal{N}, \forall j \in \mathcal{M}, \forall d \in \mathcal{K}$.
We then give user $i$'s mixed strategy aggregative contention probability function:
\begin{equation}
\label{MixProb}
Q_{ij}^d( \vec p,\vec P_d) = \gamma \sum\nolimits_{l = 1}^n {\sum\nolimits_{j = 1}^m {q_{i}^d(p_l)P_{lj}^d } }, \quad \forall d \in \mathcal{K}.
\vspace{-.1cm}
\end{equation}
where $\vec P_d$=\{\{$P_{11}^d,...,P_{n1}^d$\},..., \{$P_{1j}^d,...,P_{nj}^d$\}...,  \{$P_{1m}^d,...,P_{nm}^d$\}\} is the mixed strategy set of all users on channel $d \in \mathcal{K} $.  Noted that a user has $k$ such aggregative contention probability functions in total.
Under mixed strategy, a user's action $\pi$
 is determined as $\pi$ = $\{\phi _{ij}^{d}, p_i, P_{ij}^{d}\}, \forall i \in \mathcal{N},  \forall j \in \mathcal{M},  \forall d \in \mathcal{K}$. From the perspective of applications, we can refer each user as a ``type'' of specific application, which can be inferred
by his action profile.   Accordingly, we have following utility function for user $i$ under action $j$ and channel $d$:
\begin{equation}
\vspace{-.1cm}
U_{ij}^d(p_i,Q_{i}^d(\vec p, \vec P_d)) =\sum\nolimits_{d \in \mathcal{K}} {\log C_{ij}^d + Q_{ij}^d( \vec p,\vec P_d)}.
\end{equation}
This equation presents the utility a user $i$ can get by using channel $d$ under action $j$.  In the assumption of multiple channels and mixed strategy, we need to consider and define  the expected utility on all channels for user $i$:
\begin{equation}
\vspace{-.1cm}
U_{ij} (p_i ,Q_i^d (\vec p,\vec P)) = \mathop \mathbb{E}\nolimits_{\vec P} U_{ij}^d (p_i ,Q_i^d (\vec p,\vec P_d )).
\end{equation}

Moreover, since the high demand of spectrum limits the channel usage, we divide time into small periods. Consider a periodic mediated spectrum sharing game, three steps occur in a period as shown in Fig. 2. At the beginning of the period, users access channels with or without the suggested mixed strategies. Then, users send their mixed strategies on subsequent period's channel demands to BS. Meanwhile, BS computes a mixed strategy profile according to submitted requests. Before the end of the period, BS suggests the mixed strategy profile to each user. A user may only keep one channel for one period and have to compete for next channel usage.

In our model, there are two kinds of common target adversaries who try to learn actions of spectrum users. One adversary is the cheater who picks up information in BS. Since spectrum users submit their mixed strategies to BS and BS chooses the ideal NE for each user, it is not very hard for those cheaters to infer a user's action by the NE he achieves. The other adversary is malicious competing users who try to learn other users' actions through statistics. The fact that users conduct mixed strategy, together with our assumption that different actions contribute to different mixed strategies, makes statistics an effective way to learn users' actions. Since a user's private type can be inferred from his action, we preserve privacy of spectrum users by ensuring that a user's action is less likely to be studied.
We list important notation in this paper in \textbf{Table 1}.

\section{Aggregative Game and Privacy Model}
In this section, we review some preliminary concepts in algorithmic game theory and differential privacy. We also briefly explain how we adopt these concepts in our model.

\subsection{$\gamma$-aggregative Game}
Since a user's utility function on one channel is related to the aggregative function of every competing user's contention probability on the channel, we bring in the concept of aggregative game\cite{JENSEN}. A multi-dimension $\gamma$-aggregative game is a class of games represented by $\Gamma  = (\mathcal{N},\mathcal{M}, \mathcal{K},\{ S(\mathcal{N})_i^d\},\{ U_i^d\},\gamma)$. Specifically, $\mathcal{N}$ is the set of players, $\mathcal{M}$ is the set of actions and $\mathcal{K}$ is the dimension set of the game. Player $i$ is in the player set $\mathcal{N}$ and the aggregative function is aggregated by all users in $\mathcal{N}$. $S(\mathcal{N})_i^d$ is a player $i$'s aggregative function, or aggregator, on dimension $d$ and $\{ S(\mathcal{N})_i^d\}$ is the aggregative function set. $U_i^d$ is player $i$'s utility function on the dimension $d$ while $\{ U_i^d\}$ is the set of utility functions. $\gamma$ is the impact bound, limiting the greatest difference a player can make to the aggregative function by changing his action.


In the wireless network, the dimension in the multi-dimension $\gamma$-aggregative game is instantiated as the cardinality of the available channel set. User $i$'s aggregator on channel $d$ is the aggregative contention probability function $Q_{i}^d(\vec p)$ for the channel from his perspective. We use ``user" and ``player", ``aggregator" and ``aggregative contention probability functions" interchangeably.

 Let $Q_{i}^d(p_i',\vec p_{-i})$ be an aggregative contention probability when any single user deviates from his original action. Then, we have the following definition.
\begin{defn}
\emph{\cite{Babi} An aggregative game whose every aggregator satisfies $||Q_{i}^d(\vec p) -Q_{i}^d(p_i',\vec p_{-i})||_\infty \le \gamma$ is a $\gamma$-aggregative game.}
\end{defn}
The fact that individual impact on aggregative contention probability function is bounded by  $\gamma$ makes it a typical Lipschitz game.  Here  $\gamma$ is called Lipschitz constant of a  game \cite{AS}. Noted that this game-theoretical definition of Lipschitz is different from 1-Lipschitz mentioned above, which is a mathematical definition. We  summarize the property  of $\gamma$ in the Lipschitz game as follows.
\begin{thm}
(Lipschitz Game \cite{AS}) A game with $n$ players, $m$ actions and a Lipschitz constant $\gamma$ admits $\gamma \sqrt {{\rm{8}}n\log (2mn)}$-approximate Nash equilibria.
\end{thm}
Theorem 1 guarantees that our aggregative model in large-scale wireless network  always converges to Nash equilibria.

\subsection{Aggregative Best Response}
Aggregative best response, which is a very important concept in aggregative game \cite{Babi}, is also introduced to describe Nash equilibria in the wireless network. Let $\mathcal{P}$ be the contention probability set and $\vec p$ be the contention probability profile of all users. $p_i'$ denotes that user $i$ deviates from his original action and thus changes his contention probability. A user $i$ is playing an $\eta$-best response to $\vec p$ if his utility function satisfies $U_{ij}^d(p_i,Q_{i}^d(\vec p)) \ge \mathop {\max }\nolimits_{p_i ^\prime  } U_{ij}^d(p_i',Q_{i}^d(p_i',\vec p_{-i}))-\eta$, for all $p'_{i} \in \mathcal{P}$.
In the random access network, if all users are randomly playing an $\eta$-best response, the situation is defined as an $\eta$-approximate NE. Similarly, we define the aggregative best response $\vec {BA}_{i}(Q_{i}^d(\vec p))$ for user $i$  with aggregative contention probability $Q_{i}^d(\vec p)$ on each channel $d$  as
\begin{equation}
\label{BA3}
U_{ij}^d(p_i,Q_{i}^d(\vec p)) \ge \mathop {\max }\nolimits_{p_i ^\prime  } U_{ij}^d(p_i',Q_{i}^d(\vec p))-\eta.
\end{equation}
The aggregative best response indicates that the difference between user $i$'s worst case $\vec {BA}_{i}(Q_{i}^d(\vec p))$ and his exact aggregative best response is at most $\eta$.

\textbf{Remark 1:} Noted that a user is playing best response when his action is against other users' actions $\vec p_{-i}$, while aggregative best response is acted against aggregative contention probability function $Q_{i}^d(\vec p)$. In large-scale spectrum sharing, the effect of individual's action on the aggregative contention probability function, which is bounded by $\gamma$, is relatively small comparing to the aggregator. So in function (\ref{BA3}), we ignore the effect of the user's action on the aggregative function in the aggregative best response. The user plays aggregative best response as if the aggregative function were promised to be $Q_{i}^d(\vec p)$.

In the aggregative spectrum sharing game, we use aggregative best response instead of best response to describe a NE. In fact, an aggregative best response can translate into a best response with regard to our assumption of $\gamma$-aggregative game and the property of 1-Lipschitz, vice versa.
\begin{lem} Let $\vec p$ be the contention probability profile such that user $i$'s contention probability $p_{i}$ is an $\eta$-best response. Then, $p_{i}$
is an $(\eta + \gamma)$-aggregative best response to $U_{ij}^d(p_i,Q_{i}^d(\vec p))$.
\end{lem}
\begin{IEEEproof}
Let $Q_{i}^d(p_i',\vec p_{-i})$ be the aggregative function if any user deviates from his original action. Since $p_{i}$ is an $\eta$-best response, we know that $U_{ij}^d(p_i,Q_{i}^d(\vec p)) \ge U_{ij}^d(p_i',Q_{i}^d(p_i',\vec p_{-i}))-\eta$ . By the bounded influence of user $i$, we know that $||Q_{i}^d(\vec p) - Q_{i}^d(p_i',\vec p_{-i})||_\infty\le\gamma$. Also, combining Lipschitz property of $u_{ij}^d$ and our assumption of $\gamma$ aggregative game, we have that $|U_{ij}^d(p_i',Q_{i}^d(\vec p)) - U_{ij}^d(p_i',Q_{i}^d(p_i',\vec p_{-i})|\le\gamma$. It follows  that $ U_{ij}^d(p_i',Q_{i}^d(p_i',\vec p_{-i}) \ge U_{ij}^d(p_i',Q_{i}^d(\vec p))- \gamma$, and therefore $U_{ij}^d(p_i,Q_{i}^d(\vec p)) \ge U_{ij}^d(p_i',Q_{i}^d(\vec p))- \gamma-\eta$.
\end{IEEEproof}
If one user is playing an aggregative best response in the spectrum sharing network, we are able to know that another user is also playing an aggregative best response  as long as the aggregative contention probability functions of two users are close. We state such relation formally in the following lemma.
\begin{lem} Suppose user $i$'s contention probability $p_{i}$ is an $\eta$-aggregative best response to the aggregator function $Q_{i}^d(\vec p)$ for user $i$. Let $Q_{i}^d(p_i',\vec p_{-i})$ be the aggregative contention probability such that $||Q_{i}^d(\vec p) - Q_{i}^d(p_i',\vec p_{-i})||_\infty\le\alpha$. Then, $p_{i}$ is an $(\eta + 2\alpha)$-aggregative best response to $Q_{i}^d(p_i',\vec p_{-i})$.
\end{lem}
\begin{IEEEproof}
 Let $Q_{i}^d(p_i',\vec p_{-i})$ be the aggregative contention probability if any user deviates from his original action. Since $p_{i}$ is an $\eta$-aggregative best response to $Q_{i}^d(\vec p)$, we have $U_{ij}^d(p_i,Q_{i}^d(\vec p)) \ge U_{ij}^d(p_i',Q_{i}^d(\vec p))- \eta$. By 1-Lipschitz property and the assumption of $\gamma$-aggregative game of $U_{ij}^d(p_i,Q_{i}^d(\vec p))$, $U_{ij}^d(p_i,Q_{i}^d(p_i',\vec p_{-i})) \ge U_{ij}^d(p_i,Q_{i}^d(\vec p))- \alpha$ and also $u_{i}(a ,q) \ge u_{i}(a,q')- \alpha$.  Combining these inequalities, we have $U_{ij}^d(p_i,Q_{i}^d(p_i',\vec p_{-i})) \ge U_{ij}^d(p_i',Q_{i}^d(p_i',\vec p_{-i}))- \eta-2\alpha$.
\end{IEEEproof}
The relationship among different aggregative best responses shown in Lemma 1, together with the relationship between the aggregative best response and the best response  shown in   Lemma 2, is utilized to realize an approximate NE in the spectrum sharing. We provide the following lemma to show the connection.

\begin{lem} Let $\vec p$ be the contention probability profile such that every user is playing $\eta$-aggregative best response to $Q_{i}^d(\vec p)$. Then, we know that each user is playing ($\gamma + \eta$)-best response, and hence $\vec p$ forms a ($\gamma + \eta$)-NE.
\end{lem}
\begin{IEEEproof}
Let $Q_{i}^d(p_i',\vec p_{-i})$ be the aggregative contention probability if any user deviates from his original action. Since $p_{i}$ is an $\eta$-aggregative best response, we know $U_{ij}^d(p_i,Q_{i}^d(\vec p)) \ge U_{ij}^d(p_i',Q_{i}^d(\vec p))-\eta$. We know that $||Q_{i}^d(\vec p)- Q_{i}^d(p_i',\vec p_{-i})||_\infty\le\gamma$ by bounded influence of $i$. Then by Lipschitz property and the assumption of $\gamma$-aggregative game, $U_{ij}^d(p_i',Q_{i}^d(\vec p))\ge U_{ij}^d(p_i',Q_{i}^d(p_i',\vec p_{-i}))- \gamma$. It  follows that $U_{ij}^d(p_i,Q_{i}^d(\vec p)) \ge U_{ij}^d(p_i',Q_{i}^d(p_i',\vec p_{-i}))- \gamma-\eta$.
\end{IEEEproof}
These relations among best response, aggregative best response and NE bridge the gap between an aggregative best response and a NE. Therefore, we can achieve a NE  when all users are randomly playing aggregative best response in the spectrum sharing network.

\subsection{Mediated Game}
The introduction of a mediator transforms the incomplete information game into a complete one. In the mediated game, a user can opt-in using the mediator or neglect it. If a user opt-in, he will send his mixed channel access to the mediator. Let $P$ denote mixed strategies of opting-in users. For those who opt-out, we use notation ``$\bot$" to indicate their mixed strategies. We also define $\mathcal{A}^n$ as the output of the mediator. Intuitively, a mediator is a mechanism $ (P \cup {\rm{\{ }} \bot {\rm{\} )}}^n  \to \mathcal{A}^n$
 that takes users' mixed strategies as input,  computes a solution according to the corresponding contention probabilities and outputs a suggested mixed strategy profile for every user.


Since we assume a weak mediator in the spectrum sharing game, users have several options: they can opt-out leaving out of the BS and report $\bot$. Alternatively, they can opt-in using the BS as the mediator. Besides, opting-in users are free to decide whether to follow the suggestion or not when they receive the suggested mixed strategies.

Regardless of users' options, the mediator always gives out suggestions to all users. Therefore, there are two kinds of outputs $\mathcal{A}_i'$ for opting-in users and $\mathcal{A}_i''$ for opting-out users. $\mathcal{A}_i=\mathcal{A}_i' \cup \mathcal{A}_i''$, where
\begin{equation}
\mathcal{A}_i'=\{\mathcal{A}_i|P_i\to \mathcal{A}_i\}, \mathcal{A}_i''=\{\mathcal{A}_i|\bot\to \mathcal{A}_i,\mathcal{A}_i\quad is\quad constant\}.
\vspace{-.2cm}
\end{equation}
$\mathcal{A}_i'$ is the user-specific suggestion for opting-in users and $\mathcal{A}_i''$ is the fixed suggestion for opting-out users.

\subsection{Differential Privacy}
We then present some tools from differential privacy, which is an ideal tool to study \emph{large-scale} spectrum sharing in wireless network. Intuitively, differential privacy means that a single change in the input data set only has a limited impact on the output. This impact diminishes with the size of users growing large.
\begin{defn} \emph{(Differential Privacy \cite{DMNA}) A randomized algorithm $\mathcal{M}$ satisfies ($\epsilon,\delta$)-differential privacy if for any two input sets $A$ and $B$ with only a single input difference, and for any set of outcomes $\mathcal{R}$, we have:
\begin{equation}
\Pr(\mathcal{M}(A) \in \mathcal{R}) \le e^{\epsilon}
\Pr(\mathcal{M}(B) \in \mathcal{R}) + \delta.
\vspace{-.3cm}
\end{equation}}
\end{defn}
An adapted version of differential privacy, which is called ``joint differential privacy", is applied to suit the case of wireless network. The joint differential privacy denotes that a single change in the input data other than user $i$'s can only have a limited impact on the output to user $i$, i.e.,
\begin{defn} \emph{(Joint Differential Privacy \cite{KPRU}) Let $\mathcal{R}^{n}$ be the output profile for $n$ users. Two input sets $A$ and $B$ with size $n$ are $i$ - neighbors if they differ only in the $i$-th component. An algorithm $\mathcal{M}$ is ($\epsilon,\delta$)-joint differentially private if for any user $i$, for any pair of $i$-neighbors input $A$ and $B$, and for any subset of outputs $S\subseteq \mathcal{R}^{n-1}$,
\begin{equation}
\Pr(\mathcal{M}(A)_{-i}) \in \mathcal{S}) \le e^{\epsilon}
\Pr(\mathcal{M}(B)_{-i}) \in \mathcal{S}) + \delta.
\vspace{-.4cm}
\end{equation}}
\end{defn}

One important property of differential privacy is the adaptive composition theorem. In the proposed online learning algorithm,  BS needs as many as $T$ periods to learn users' strategies. This makes the situation $T$-fold adaptive.
\begin{thm} (Adaptive Composition \cite{DRV}) Let $\mathcal{A}$ be the input set and mechanism be $\mathcal{M}: \mathcal{A}^{n} \to \mathcal{R}^{T}$ in $T$ periods. If $\mathcal{M}$ is an adaptive composition of as many as $T$ individual ($\epsilon,\delta$)-differentially private mechanisms, then $\mathcal{M}$ satisfies ($\epsilon',T\delta+\delta'$)-differential privacy for
\begin{equation}
\vspace{-.4cm}
\epsilon' = \epsilon\sqrt{2Tln(1/\delta')} + T\epsilon(e^{\epsilon} - 1).
\end{equation}
\end{thm}

An immediate and useful  lemma provides us with a ($\epsilon,\delta$)-differentially private mechanism.
\begin{lem}
For any $\epsilon \le 1$ and $\delta \ge 0$ , if $M$ is a $T$-fold adaptive composition of ($\epsilon/\sqrt{8Tln(1/\delta)}$, 0)-differentially private mechanisms, then $M$ satisfies ($\epsilon,\delta$)-differential privacy.
\end{lem}

\subsection{Truthfulness}
Intuitively, a mechanism is incentive compatible, or called  ``truthful", when players can get the highest utility only by acting truthfully. In other words, truth telling is one's dominant strategy.

\begin{defn} \emph{(Incentive Compatible \cite{AGT}) A mechanism is called incentive compatible (IC) if every user can achieve the best outcome just by telling the true information.}
\end{defn}

We slightly modify this definition by assuming that a mechanism is incentive compatible if the user only has the incentive to opt-in rather than opt-out.

\subsection{$\gamma$-approximate \emph{Ex-post} NE}
Finally, we provide the definition of $\gamma$-approximate \emph{ex-post} NE, which is the equilibrium we concern in this paper.
Intuitively in game theory, NE is a solution concept in which no player can get a better payoff by unilaterally changing his strategy or action.


Compared to the standard NE, our $\gamma$-approximate\emph{ ex-post} NE has two major differences. The first one is that an \emph{ex-post} NE is achieved without the knowledge of priori distribution of strategies. The second one is that the $\gamma$-approximate NE reaches an equilibrium solution that  deviates from the exact NE by most $\gamma$.

\begin{defn} \emph{($\gamma$-approximate \emph{ex-post} NE \cite{MS}) Assuming a collection of strategies $\{ \sigma _i :\mathcal{P} \to \mathcal{A}_i \} _{i = 1}^n$.
 A $\gamma$-approximate \emph{ex-post} NE is formed if for any mixed strategy in $
 p\in \mathcal{P}^n$, and for any player $i$ and action $a_i \in \{\mathcal{A}_i\}$:
\begin{equation}
\vspace{-.4cm}
 u_i ^\prime  (\sigma _i (p_i ),\sigma _{ - i} (p_{ - i} )) \ge u_i ^\prime  (a_i ,\sigma _{ - i} (p_{ - i} )) - \eta.
\end{equation}}
\end{defn}

Due to the fact that users are not able to know the prior distribution of strategies, the \emph{ex-post} NE is a very strong solution concept for incomplete information games. In the following two sections, we detail private and truthful equilibrium computation carried out in the mediator.


\section{Mediated Private and Truthful Game}
\subsection{Searching for an Ideal Equilibrium}
In a mediated spectrum sharing game, if all non-cooperative users conduct pure strategy to compete for channels, pure strategy Nash equilibria exist. In terms of the best response, a network performs $\eta$-approximate NE when all users are randomly playing $\eta$-best responses in contention probability profile $\vec p$ under pure strategy.  However, real world selfish users conduct mixed channel access strategies to get better utility. With mixed strategy, whether a mixed strategy NE exists in the network is unknown. Consequently, designing new computation schemes to search for an ideal NE is essential.

 Since we focus on the aggregative mediated spectrum sharing, users send their mixed channel access strategies ${\vec P}   = ({\vec P}_1  ,...,\vec {P_i},...,{\vec P}_n  )$, where  $\vec {P_i}= (P_{ij}^1,...,P_{ij}^k) $ to the BS. With these submitted channel access strategies, we need to find the aggregative contention probability for every user so that they can randomly achieve $\xi$-aggregative best response: $\xi \!\!- \!\!\vec {BA}_{i}(Q_{i}^d(\vec p, \vec P_d))$ with mixed strategy. As mentioned earlier, we reach aggregative best response through best response.


We first define the ideal aggregative contention probability $\widehat Q_{ij}^d( \vec p,\vec P_d)$ as the optimal set-valued aggregative function for each user when he accesses the channel $d$. To find the set-valued aggregative contention probability function for each user, we round the whole range of aggregative contention probability $[ 0, n\gamma] ^k$ for all $k$ channels.
In the process, we  search through the discretized grid of the whole possible space $\{ 0,\alpha,..., n\gamma\} ^k$ of aggregative function for all users as well as all channels. Since there are several equilibria for user $i$ on channel $d$ in the assumption of pure strategy, all the corresponding set-valued aggregative functions can be found. However, with our goal to provide an efficient mechanism, we need to select the set-valued aggregative function $\widehat Q_{ij}^d( \vec p,\vec P_d)$ that has the minimum difference $a(\widehat Q_{ij}^d( \vec p,\vec P_d))$ with users' aggregative probability function on that channel. See the objective function in (\ref{eq:Apart1}) and constraints (\ref{equcross1}) and (\ref{equcross2}). This difference is the cost brought by the mixed strategy. Noted that the mixed strategies submitted to the BS for who opt-out are considered $\frac{1}{k}$ in (\ref{equcross1}) and (\ref{equcross2}).  With the assumption of 1-Lipschitz condition, the minimum difference between two aggregative functions leads to the minimum difference between the corresponding utilities.

Noted that our model is a general model of spectrum sharing network, which contains models of spectrum reuse  \cite{CH} and conflicting graph \cite{ZS}. On the one hand,  a user will allocate  probability ``0" to those channels in $\mathcal{K}$ that are too far away for transmission. Obviously, faraway channels can only be utilized by its surrounding users.  On the other hand,  if two interfering (or conflicting) channels are in the transmission range of a user, we can only assign a probability
 to one channel and set the other one to be zero, and set the rest of non-zero mixed probabilities on other  accessible non-conflicting channels. However, we do not plan to go into details for the issues of spatial reuse and channel conflict in this paper. In short, if a user cannot play aggregative best response on channel $d$, his mixed strategy on it will be zero.

Summarizing above discussions, we can define following formulation for each user $i$:
\begin{equation}
\vspace{-.2cm}
\min \quad a(\widehat Q_{ij}^d( \vec p,\vec P_d)),\IEEEyesnumber \label{eq:Apart1}
\end{equation}
\begin{equation}
\vspace{-.2cm}
\label{equcross1}
\gamma \sum\limits_{l = 1}^n {\sum\limits_{j = 1}^m {q_{i}^d(p_l)P_{lj}^d } }  \le  \widehat Q_{ij}^d( \vec p,\vec P_d) + a(\widehat Q_{ij}^d( \vec p,\vec P_d)),\quad\forall d \in \mathcal{K},
\end{equation}
\begin{equation}
\vspace{-.2cm}
\label{equcross2}
\gamma \sum\limits_{l = 1}^n {\sum\limits_{j = 1}^m {q_{i}^d(p_l)P_{lj}^d } }  \ge  \widehat Q_{ij}^d( \vec p,\vec P_d) - a(\widehat Q_{ij}^d( \vec p,\vec P_d)),\quad\forall d \in \mathcal{K},
\end{equation}
\begin{equation}
\vspace{-.2cm}
\label{equaindi}
0 \le P_{ij}^d  \le 1,\quad \forall i\in\mathcal{N}, \quad j \in \xi\!-\!\vec {BA} _i(\widehat Q_{ij}^d( \vec p,\vec P_d)),
\end{equation}
\begin{equation}
\vspace{-.2cm}
\label{BA1}
P_{ij}^d  = 0,\quad\forall i\in\mathcal{N},\quad j \notin \xi\!-\! \vec {BA} _i(\widehat Q_{ij}^d( \vec p,\vec P_d)),
\end{equation}
\begin{equation}
\vspace{-.2cm}
\label{BA2}
\sum\limits_{d = 1}^k {P_{ij}^d }  = 1,\quad\forall i\in\mathcal{N},j\in \mathcal{M}.
\end{equation}
where $\xi=\gamma+2\alpha+\zeta$. $\xi$ is the result of lemma 1, 2 and 3. Noted that $\zeta \ge
\gamma \sqrt {{\rm{8}}n\log (2mn)}$, which is introduced on account that the Lipschitz game admits $\gamma \sqrt {{\rm{8}}n\log (2mn)}$-approximate Nash equilibria. $\gamma$ comes from the assumption of $\gamma$-aggregative game. According to lemma 1, 2, we add up these two parameters. Since we discretize the aggregative contention probability space by $\alpha$, $2\alpha$ is added to compensate the gap among discretized aggregative functions due to lemma 3.



\subsection{An Efficient Private-Preserving  Equilibrium Selection Algorithm}

Now, we show how to devise an efficient privacy-preserving  equilibrium selection algorithm. With our goal to find the minimum difference $a$, we need to compare every possible value of contention probability function to the aggregative contention probability functions constituted by users' submitted mixed strategies. Since there are as many candidates as $
\left( {\frac{{n\gamma }}{\alpha }} \right)^k$ to check, together with our goal to prevent cheating learning in the BS, the design of efficient privacy-preserving algorithm can be very challenging.

However, since we only need to output one contention probability function on one channel for each user, we introduce the \emph{sparse vector mechanism} to deal with such large aggregative contention probability candidates data flows. The mechanism is originated from differential privacy \cite{DR}. To fit it in the aggregative spectrum sharing game, we devise a similar mechanism and name it the Sparse Cost for Aggregative Contention Probability (\textbf{SparCost}),  shown in Algorithm 1.

To find the lowest cost, our algorithm first takes in a sequence of costs $a(\widehat Q_{ij}^d( \vec p,\vec P_d))$ for aggregative contention probability functions and a cost threshold $T$ which is chosen according to past channel usage. The cost threshold is  used as a benchmark in our algorithm. Then, the \textbf{SparCost} adds noise subject to Laplace distribution to the cost threshold $T$ and the cost $a(\widehat Q_{ij}^d( \vec p,\vec P_d))$ respectively. Noted that adding noises to two inputs is essential to protect user's mixed strategy as well as their types. The privacy
 concern brings about an error bound $e_1$, which we will explain in details in following Theorem 3. Next, we compare the noisy cost ${{\mathord{\buildrel{\lower3pt\hbox{$\scriptscriptstyle\frown$}} \over a_n(Q_i^d)} }}$ with the noisy cost threshold ${{\mathord{\buildrel{\lower3pt\hbox{$\scriptscriptstyle\frown$}} \over T} }}$. By comparing the input costs and the threshold, we can find a cost that is fairly small and thus obtain the corresponding aggregative contention probability as well as the ideal equilibrium. Finally, the \textbf{SparCost} outputs the first noisy cost that is below the noisy threshold.

The corresponding aggregative function is the set-valued aggregative contention probability we are looking for. Before the ideal cost is found, the \textbf{SparCost} reports $\bot$ for all costs that are above the threshold. Noted that the algorithm runs at most $k$ times for each user and so runs $nk$ times in total. In the algorithm, $v_n$ means the $n$-th noise add to the $n$-th cost when every user needs to check $N=\frac{n\gamma}{\alpha}$  costs for every channel.

\begin{algorithm}[!htb]
 \caption{Sparse Cost for Aggregative Contention Probability \textbf{SparCost}($\{a(\widehat Q_{ij}^d( \vec p,\vec P_d))\}$,$T$,$1$,$\varepsilon$)}  
 \label{alg:example1}  
 \KwData{an adaptively chosen stream of costs $a(\widehat Q_{ij}^d( \vec p,\vec P_d))$ for set-valued aggregative contention probability function $\widehat Q_{ij}^d( \vec p,\vec P_d)$ of sensitivity $\gamma$, a threshold $T$, total number of throughput answers 1, and privacy parameter $\varepsilon$.}
 \KwResult{An answer $\{ {\theta_i}\}$ for user $i$. }

  \textbf{Let} $\mathord{\buildrel{\lower3pt\hbox{$\scriptscriptstyle\frown$}}\over T}  = T + Lap(\frac{{2\gamma }}{\varepsilon })$.

 \textbf{let} $\sigma  = \frac{{4\gamma }}{\varepsilon }$.


 \For{ $a(\widehat Q_{ij}^d( \vec p,\vec P_d))$  }{

  \textbf{Let} ${v_n} = Lap(\sigma )$ and ${{\mathord{\buildrel{\lower3pt\hbox{$\scriptscriptstyle\frown$}} \over a_n(Q_i^d)} }} = a(\widehat Q_{ij}^d( \vec p,\vec P_d)) + {v_n}$.

  \If{${{\mathord{\buildrel{\lower3pt\hbox{$\scriptscriptstyle\frown$}} \over a_n(Q_i^d)} }} \le \mathord{\buildrel{\lower3pt\hbox{$\scriptscriptstyle\frown$}} \over T} $}{
  \KwOut{${\theta_i} = {{\mathord{\buildrel{\lower3pt\hbox{$\scriptscriptstyle\frown$}} \over a_n(Q_i^d)} }} .$}
%
   }
   \Else
   {
   \KwOut{$ \bot $.}
  }
 }
\end{algorithm}

\begin{thm} Let $e_1$ be the error bound for the output aggregative function. For any sequence of $N$ costs $a_1(\widehat Q_{ij}^d( \vec p,\vec P_d)$), ..., $a_N(\widehat Q_{ij}^d( \vec p,\vec P_d))$ for user $i$ on channel $d$ such that$|n: a_n(\widehat Q_{ij}^d( \vec p,\vec P_d)) \le T +e_1| \le c$, \textbf{SparCost} satisfies $\varepsilon$-differential privacy and, with probability at least $1-\frac{\beta}{2}$ releases cost answers such that $|\theta_n  - a_n(\widehat Q_{ij}^d( \vec p,\vec P_d))| \le e_1$,
 and for all $\theta_n$=$ \bot $, $a_n(\widehat Q_{ij}^d( \vec p,\vec P_d)) \ge T - e_1,$ where
\begin{equation}
e_1  = \frac{{8\gamma (\log N + \log (4/\beta )}}{\varepsilon }.
\end{equation}
 In the \textbf{SparCost}, $N$ is $
\left( {\frac{{n\gamma }}{\alpha }} \right)$ for one user on one channel. Since a user has $k$ channels to choose from , each channel can bring an $e_1$. Also, for different users, their aggregative contention functions have different $e_1$. For simplicity, we take the biggest of them and give the overall error bound brought by \textbf{SparCost}:
\begin{equation}
E_1= \max _{i\ \to e_1 } \frac{{8\gamma (k\log \frac{{n\gamma }}{\alpha } + \log (4/\beta ))}}{\varepsilon }.
\end{equation}
Combining the approximation brought by discretion, the total approximation brought by \textbf{SparCost} is $\alpha  + E_1$.
\end{thm}
The proof is given in Appendix A.

The algorithm \textbf{SparCost} guarantees that the output cost and the corresponding aggregative contention probability are differentially private and have low sensitivity to users' submitted strategies. Hence, any malicious user or opting-out user can only have little effect on the search for ideal aggregative contention probability function. Besides, \textbf{SparCost} protects the privacy of individual mixed strategy,  which constitutes the aggregative contention probability function. It prevents cheating learning in BS. Therefore, the algorithm provides us with privacy guarantee.



As illustrated in Fig. 3, there are different Nash equilibria with different amount of users. The left part of Fig. 3 shows that aggregative contention probabilities of freely competing users can hardly get close to any of set-valued aggregative functions. Meanwhile, the right side demonstrates the performance of \textbf{SparCost}, showing that resulting aggregative contention probabilities are much more closer to those set-valued aggregative functions for constraint (\ref{equcross1}).
Even though we have obtained the expected aggregative contention probability for each user, we aim to give out a strategy profile solution so that all users can get their optimal utilities. In the next section, we propose the learning algorithm that fulfills our goal to provide a private and truthful NE solution for all users.

\section{Private and Truthful NE Computation}

In this section, we discuss how to compute the strategy profile solution via an online learning algorithm. We also show that the solution achieves a NE and has better performance when the spectrum sharing is in large scale. The solution guarantees privacy by satisfying joint differential privacy and ensures truthfulness.
\begin{figure}
\centering
\includegraphics[scale=.55]{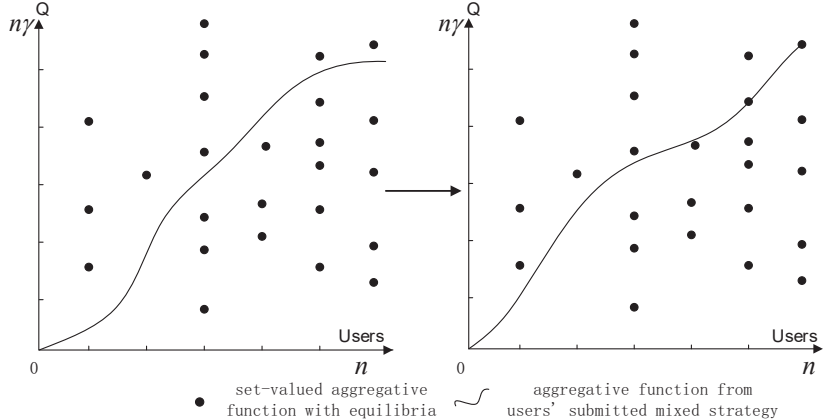}
\caption{\textbf{SparCost}'s performance on searching for objective aggregative function from original aggregative function}
\label{fig:digraph}
\vspace{-.2cm}
\end{figure}

\subsection{Rounded Exponential Mechanism  for Preserving Privacy}

Although \textbf{SparCost} guarantees privacy of the users' submitted mixed strategies, it is not enough to ensure that the output strategy solution has low sensitivity to the input submitted mixed strategy. We introduce differential privacy via \emph{Exponential Mechanism} to induce joint differential privacy to the output strategy solution.


\begin{defn}(Exponential Mechanism \cite{DR}) Let $f(x,r)$ be a function of the input $x$ and the result of the mechanism $r$. The exponential mechanism $\emph{EXP}(x, f(x, r), \varepsilon)$ selects and outputs an element $r$ with probability proportional to
$\exp (\frac{{\varepsilon f(x,r)}}{{2\Delta f}})$.
\end{defn}

The exponential mechanism has following property:
\begin{thm} \cite{DR} \emph{EXP}(x, f, $\varepsilon$) satisfies $\varepsilon$-differential privacy and, with probability at least 1-$\beta$, outputs an outcome r such that
\begin{equation}
\vspace{-.2cm}
f(x,r) \ge \mathop {\max }\nolimits_{r'} f(x,r') - \frac{{2\Delta (f)(\log |R|/\beta )}}{\varepsilon },
\end{equation}
where $|R|$ denotes the output space range,  which is $\frac{n\gamma}{\alpha}$ in our case.

\end{thm}

To satisfy the cross constraint (\ref{equcross1}) well partitioned among all $n$ users , we regard the cross constraint as the score function $f$ in the \emph{EXP} for each user $i$ under channel $d$:
\begin{equation}
\vspace{-.1cm}
f(\vec P_d, Q_{ij}^d( \vec p,\vec P_d)) = \gamma \sum\limits_{l = 1}^n {\sum\limits_{j = 1}^m {q_{i}^d(p_l)P_{lj}^d } }  - \lambda_d,  \forall d \in \mathcal{K}, \forall i \in \mathcal{N}.
\end{equation}
 The parameter $\lambda_d$, which equals to $\widehat Q_{ij}^d( \vec p,\vec P_d) + \alpha + E_1$, is the upper difference bound between user $i$'s aggregative contention probability on channel $d$ and the set-valued aggregative function. Noted that $\widehat Q_{ij}^d( \vec p,\vec P_d)$ is obtained from \textbf{SparCost} and $a$ is replaced by $\alpha + E_1$.

Every user $i$ has as many as $k$ heterogeneous impacts on the aggregative contention probability and so possesses $k$ set-valued aggregative contention probabilities. We implement above exponential mechanism in   $k$ rounds for each user and devise the following Algorithm 2 of Rounded Exponential Mechanism (REM) for preserving privacy. We name it \textbf{REXP}.



%


 The Algorithm 2 proceeds in rounds and we only provide one round here. For each user, BS feeds mixed strategies ${\vec P}$ to \emph{EXP}$({\vec P} ,f,\varepsilon)$ and selects an aggregative contention probability for each channel respectively.
Then, the \emph{EXP} outputs these $k$ selected aggregative contention probability functions $\{Q_{ij}^d( \vec p,\vec P_d), \forall d\in \mathcal{K} \}=\{Q_{ij}^1( \vec p,\vec P_1),...,\{Q_{ij}^d( \vec p,\vec P_d)\},...,\{Q_{ij}^k( \vec p,\vec P_k)\}\}$ for the user. For user $i$, his individual contention probabilities $q_{i}^d(p_i)$=$\frac{{\log p_{i}}}{{\log p_{i} + \sum\nolimits_{l \ne i} {\log (1 - p_l )} }}$ from selected functions are the key to update  his own part of mixed strategy suggestion.
Such update is carried out by following online learning algorithm.

\begin{algorithm}[!htb]
 \caption{REM for preserving privacy: \textbf{REXP}($\{\widehat Q_{ij}^d( \vec p,\vec P_d)\}$, \emph{EXP}(${\vec P} $, $f(\vec P_d, Q_{ij}^d( \vec p,\vec P_d))$, $\varepsilon_0$), $\delta$,$\beta$)}  
 \label{alg:example2}  
 \KwData{set-valued aggregative contention probability set \{$\widehat Q_{ij}^d( \vec p,\vec P_d)\}$ for user $i$, the corresponding $\lambda_d$ on channel $d$ in \emph{EXP}, strategy variables for all users ${\vec P} $ , the score function $f$, privacy parameters ($\varepsilon_0$, $\delta$) and confidence parameter $\beta$.}
 \KwResult{An aggregative contention probability profile $\{Q_{ij}^d( \vec p,\vec P_d)\}$ satisfying $\varepsilon$-differential privacy.}
 \textbf{Initialize} ${\vec P}   = ({\vec P}_1  ,...,{\vec P_n}  )$.

\textbf{Let} $\varepsilon=\varepsilon_0\sqrt{8Tln(1/\delta)}$.

 \For{each channel $d$ and each user $i$}{

      \textbf{Let} $\Pr (Q_{ij}^d( \vec p,\vec P_d)) = \frac{{\exp (\varepsilon f(\vec P_d, Q_{ij}^d( \vec p,\vec P_d))}}{{\sum\nolimits_{d \in k} {\exp (\varepsilon f(\vec P_d,Q_{ij}^d( \vec p,\vec P_d))} }}.$

       Select $Q_{ij}^d( \vec p,\vec P_d)$ according to the high probability bound $\frac{\beta}{2}$.

  }

  \KwOut{$Q_{ij}^d( \vec p,\vec P_d)$.}

\end{algorithm}

\begin{thm} The Rounded Exponential Mechanism for preserving privacy described in Algorithm 2 satisfies ($\varepsilon$, $\delta$)-differential privacy.
\end{thm}

\begin{IEEEproof}
We first prove that \emph{EXP}(${\vec P}$, $f(\vec P_d, Q_{ij}^d( \vec p,\vec P_d))$, $\varepsilon_0$) is $\varepsilon_0$ differentially private. For each user, there are at most $k$ aggregative contention probabilities. For user $i$, \emph{EXP} outputs some aggregative contention probabilities $Q_{ij}^d( \vec p,\vec P_d)$ on two neighboring inputs ${\vec P}$ and ${\vec P}'$. For users that form ${\vec P}$, if one of them changes his mixed strategy, the resulting mixed strategy distribution forms ${\vec P}'$. We use $Q_{ij}^d$ for simplicity of expression.
$$ \frac{Pr[\emph{EXP}({\vec P}, f(\vec P_d, Q_{ij}^d), \varepsilon_0)=Q_{ij}^d]}{Pr[\emph{EXP}({\vec P}', f(\vec P_d', Q_{ij}^d), \varepsilon_0)=Q_{ij}^d]}
 = \frac{\big(\frac{exp(\frac{{\varepsilon_0} f(\vec P_d, Q_{ij}^d)}{2{\Delta}f})}{\sum_{d \in \mathcal{K}} exp(\frac{{\varepsilon_0} f(\vec P_d, Q_{ij}^d)}{2{\Delta}f})}\big)}{\big(\frac{exp(\frac{{\varepsilon_0} f(\vec P_d', Q_{ij}^d)}{2{\Delta}f})}{\sum_{d \in \mathcal{K}} exp(\frac{{\varepsilon_0} f(\vec P_d', Q_{ij}^d)}{2{\Delta}f})}\big)}
= \big( \frac{exp(\frac{{\varepsilon_0} f(\vec P_d, Q_{ij}^d)}{2{\Delta}f})}{exp(\frac{{\varepsilon_0} f(\vec P_d', Q_{ij}^d)}{2{\Delta}f})}) \cdot \big( \frac{\sum_{d \in \mathcal{K}} exp(\frac{{\varepsilon_0}  f(\vec P_d', Q_{ij}^d)}{2{\Delta}f})}{\sum_{d \in \mathcal{K}} exp(\frac{{\varepsilon_0} f(\vec P_d, Q_{ij}^d)}{2{\Delta}f})}\big)$$
$$\hspace{3.5cm}= exp( \frac{\varepsilon_0 (f(\vec P_d, Q_{ij}^d)- f(\vec P_d', Q_{ij}^d))}{2{\Delta}u} )
\cdot \big( \frac{\sum_{d \in \mathcal{K}} exp(\frac{{\varepsilon_0}  f(\vec P_d', Q_{ij}^d)}{2{\Delta}f})}{\sum_{d \in \mathcal{K}} exp(\frac{{\varepsilon_0} f(\vec P_d, Q_{ij}^d)}{2{\Delta}f})}\big)$$
$$\hspace{1.2cm}\leq exp(\frac{\varepsilon_0}{2}) \cdot exp(\frac{\varepsilon_0}{2}) \cdot \big( \frac{\sum_{d \in \mathcal{K}} exp(\frac{{\varepsilon_0}  f(\vec P_d', Q_{ij}^d)}{2{\Delta}f})}{\sum_{d \in \mathcal{K}} exp(\frac{{\varepsilon_0} f(\vec P_d, Q_{ij}^d)}{2{\Delta}f})}\big)$$
$$\hspace{-4.3cm}= exp(\varepsilon_0).$$

Similarly, $ \frac{Pr[\emph{EXP}({\vec P}', f(\vec P_d, Q_{ij}^d), \varepsilon_0)=Q_{ij}^d]}{Pr[\emph{EXP}({\vec P}, f(\vec P_d', Q_{ij}^d), \varepsilon_0)=Q_{ij}^d]} \geq exp(-\varepsilon_0)$ by symmetry.
Therefore, \emph{EXP}(${\vec P}$, $f(\vec P_d, Q_{ij}^d( \vec p,\vec P_d))$, $\varepsilon_0$) satisfies $\varepsilon_0$ differential privacy. Since the following over time learning goes through $T$ periods, \textbf{REXP} restarts every $T$ periods. It satisfies ($\varepsilon$, $\delta$)-differential privacy when we set $\varepsilon=\varepsilon_0\sqrt{8Tln(1/\delta)}$ according to Theorem 2 of the adaptive composition and the corresponding lemma 4.
\end{IEEEproof}

We have got the aggregative contention probability $Q_{ij}^d( \vec p,\vec P_d)$ on each channel for each user and proved that the aggregative contention probability is ($\varepsilon$, $\delta$)-differentially private. Then, we show how to compute the private channel access strategy solution via an online learning algorithm.

\begin{figure}
\centering
\includegraphics[scale=.65]{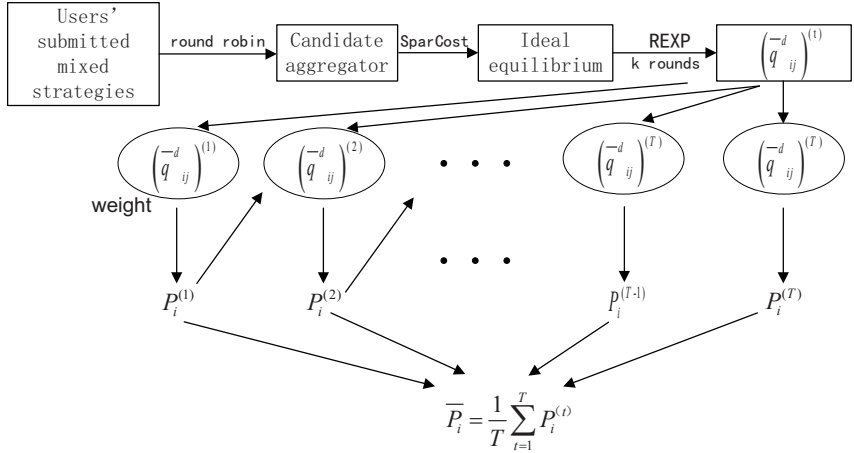}
\caption{Online learning procedure for each user}
\label{fig:digraph}
\vspace{-.2cm}
\end{figure}

\subsection{Online Learning: Performance of our result}


Due to the speculation of selfish users, the efficiency of NE based on one or a few sets of users' mixed strategies varies. The performance of one time slot NE may suffer. In large-scale spectrum sharing, users need to repeatedly make channel choice decisions. Therefore, we utilize these past accumulated information of mixed channel access strategies to improve the efficiency of the approximate NE.

The key idea of online learning is to adapt a user's mixed strategy decision based on its accumulated experiences. We achieve our goal to suggest a mixed channel access strategy profile for all users by maintaining weights on former mixed strategy experience. We here assume that our online learning algorithm refresh itself every $T$ periods. Using a doubling trick, we provide a specific period number $T$ to show the optimal result. Similar to \cite{ARORA}, we name our algorithm Multiplicative Weights for strategy online learning, or \textbf{MW}.

To implement \textbf{MW}, we need to define some notations at first. For each period $t\in\{1,...,T\}$, let ${\vec P}^t   = ({\vec P}_1^t  ,...,{\vec P}_n^t  )$ be the overall mixed strategy profile of all users at period $t$,  while $\vec {P_i^t}= (P_{ij}^1,...,P_{ij}^k) $ is the individual
mixed channel access strategy over all channels at period $t$. Also, we use $\overline q_{i}^d(p_i)=\{ q_{i}^1(p_i),..., q_{i}^k(p_i)\}$ to represent user $i$'s normalized channel connection probability on all $k$ channels.

At the first initialization period, our algorithm initializes mixed strategies of all users ${\vec P}^1$ by $\frac{1}{k}$ and feed them into \textbf{REXP}. Then for each user $i$, \textbf{REXP} processes these mixed strategies and output $k$ aggregative contention probabilities $Q_{ij}^d( \vec p,\vec P_d))$ for $k$ channels. Next, the user's weight on the mixed strategy for each channel is updated by multiplying them with his normalized channel connection probabilities $\overline q_{i}^d(p_i)$ from the output of \textbf{REXP}. And then mixed strategy is updated. Next, our algorithm projects user $i$'s updated mixed strategy into the $P_{ij}^d$, which is from his submitted individual constraint set (\ref{equaindi}). The submitted individual set is associated with the user's \emph{type} and represented by his mixed strategy. Noted that for those who opt-out, their submitted mixed strategies are considered $\frac{1}{k}$ for each channel in their individual sets. After the update, the first period is then finished.

In the following period, the updated mixed strategy from last period is taken as the input to \textbf{REXP}, and other procedures remain the same. Since there are $T$ successive periods, the weight for each user $i$, as well as the corresponding mixed strategy, is updated by $T$ times. Finally, the algorithm averages the mixed strategies for user $i$ at each period and suggests it to the user. Intuitively, with higher weight on higher connection probability, this learning method works out better in the long run. The whole process is illustrated in Fig. 4. Noted that algorithm \textbf{REXP} needs the set-valued aggregative contention probability from \textbf{SparCost}, which is the results of users' submitted mixed strategies.

\begin{algorithm}[!htb]
 \caption{Multiplicative Weights for strategy online learning:  \textbf{MW}( $\{\widehat Q_{ij}^d( \vec p,\vec P_d)\}$, $E_2$, $\beta$, $\varepsilon_0$, $\delta$)}  
 \label{alg:example2}  

 \KwData{The set-valued contention probability set \{$\widehat Q_{ij}^d( \vec p,\vec P_d)$\} for all users at each period, accuracy parameter $E_2$, and confidence parameter $\beta$, privacy parameters ($\varepsilon_0$, $\delta$). }

 \KwResult{A solution $\overline P$=\{$\overline P_1,...,\overline P_n$\}.}

 \textbf{Initialize} ${P^1}$: $P_{ij}^d = \frac{1}{k}$ for all $i\in\mathcal{N}$ and all $d\in\mathcal{K}$.

Initialize number of periods $T$ and update parameter $\eta$.

 \textbf{Let} $T = \frac{{16{n^2}{\gamma ^2}\log k}}{{{E_2 ^2}}},$  $\eta  = \frac{E_2}{4n\gamma}.$

 \For{each period $t \in \{ 1,...,T\}$ and each user $i\in \mathcal{N}$}{


 $\{Q_{ij}^d( \vec p,\vec P_d)\}$=\textbf{REXP}($\{\widehat Q_{ij}^d( \vec p,\vec P_d)\}$, $EXP({\vec P}^t $, $f(\vec P_d^t, Q_{ij}^d( \vec p,\vec P_d^t))$,$\varepsilon_0$), $\delta$, $\beta$).

 Then, we have the corresponding channel connection probability
  $q_{i}^d(p_i)$=$\frac{{\log p_{i}}}{{\log p_{i} + \sum\nolimits_{l \ne i} {\log (1 - p_l )} }}$ for each channel $d$.

  \textbf{for}
  \[\mathord{\buildrel{\lower3pt\hbox{$\scriptscriptstyle\frown$}}\over P} _{i}^{t + 1} =P_{i}^t\cdot \exp ( - \eta  \cdot \overline {q}_{i}^d(p_i)). \]

  \textbf{Projection} with relative entropy:
  \[P_i^{t + 1} = \textbf{RE}(\vec {P_i^t}||\mathord{\buildrel{\lower3pt\hbox{$\scriptscriptstyle\frown$}}
  \over P} _{i}^{t + 1}).\]

  \KwOut{the suggestion strategy for user $i$  $\overline P_i  = \frac{1}{T}\sum\nolimits_{t = 1}^T {{P_i^t}}.$}
 }

\end{algorithm}

The output of \textbf{MW} forms the suggestion strategy distribution for all users. With regard to such distribution, we present user $i$'s aggregative contention probability function on each channel $d$:
\begin{equation}
\vspace{-.1cm}
Q_{ij}^d( \vec p,\overline P_d)) =\gamma \sum\nolimits_{l = 1}^n {q_{i}^d(p_l)P_{lj}^{d}} .
 \end{equation}
In our result, one's mixed strategy is no longer specifically associated with action $j$. A user conducts the suggested mixed strategy only to satisfy aggregative best response $\xi\!-\!\vec {BA} _i(\widehat Q_{ij}^d( \vec p,\vec P_d))$ (constraints (\ref{BA1}) and (\ref{BA2})). Here $\overline P_d$ denotes each user's suggested mixed strategies on channel $d$. Therefore, the corresponding throughput, or utility, is

\begin{equation}
\vspace{-.2cm}
U_{ij}^d(p_i,Q_{i}^d(\vec p, \overline P_d)) = \log C_{ij}^d + Q_{ij}^d( \vec p,\overline P_d)= \log B_{i}^{d} \log _2 (1 + \frac{{\phi _{ij}^d \emph{g}^d_{i} }}{{\omega _{i}^d }}) + \gamma \sum\nolimits_{l = 1}^n {q_{i}^d(p_l)P_{lj}^{d}}.
\end{equation}
For each user $i$, his expected utility is then:
\begin{equation}
\vspace{-.2cm}
U_{ij} (p_i ,Q_i^d (\vec p,\overline P)) = \mathop \mathbb{E}\nolimits_{\overline P} U_{ij}^d (p_i ,Q_i^d (\vec p,\overline P_d )).
\end{equation}

\begin{lem}(billboard lemma\cite{HHRRW}). Suppose $
M:X^n  \to R$ is ($\varepsilon$,$\delta$)-differentially private. Consider any set of functions $
F_i  = X_i  \times R \to R'
$, where $X_i$ is user $i$'s input data. The composition \{$F_i (\Pi _i X,M(X))$\} is ($\varepsilon$,$\delta$)-joint differentially private, where $\Pi _i$ is the projection to user $i$'s data.
\end{lem}

The billboard lemma enables us to achieve joint differential privacy. We then have following theorem:
\begin{thm} \textbf{MW}( $\{\widehat Q_{ij}^d( \vec p,\vec P_d)\}$, $E_2$, $\beta$, $\varepsilon_0$, $\delta$) satisfies ($\varepsilon$,$\delta$)-joint differential privacy.
\end{thm}
\begin{IEEEproof}
In particular, the algorithm gets aggregative contention probability on certain channel $d$ for each user at each period via \textbf{REXP}, which satisfies ($\varepsilon$,$\delta$)-differentially privacy. All these aggregative contention probabilities form the billboard where each user can know his own impact on others. Each user only utilizes his connection probability on each channel from his aggregative contention probability to update his mixed strategy on the corresponding channel. And then, his updated mixed strategy is projected to his constraint set. With regard to the \emph{billboard lemma}, it is safe to conclude that \textbf{MW}  satisfies ($\varepsilon$,$\delta$)-joint differential privacy.
\end{IEEEproof}

Since the output strategy solution is ($\varepsilon$,$\delta$)-joint differentially private, a single change in the input data other than user $i$'s can only have a limited impact on the output to user $i$, which prevents deliberate coalition from malicious users. Competing users who try to learn actions of other users can hardly make any inference through statistics.

We then show that our computation solution is effective. In particular, our result achieves $\eta$-approximate \emph{ex-post} NE in the multi-channel wireless network, where $\eta$ is the approximation brought by the computation. The large-scale spectrum sharing and the fear of leaking types prevent users from getting the knowledge of others' strategies, giving rise to the absence of priori distribution. This makes the situation \emph{ex-post}. Recall the verdict of Lipschitz game, a Lipschitz game always converges to approximate Nash equilibria. Consequently, we conclude that our computation satisfies $\eta$-approximate \emph{ex-post} NE.

\begin{thm}
Let $\zeta  \ge \gamma \sqrt {{\rm{8}}n\log (2mn)}$, $
\varepsilon$, $\delta$, $\beta$ $\in$ (0,1). Our mechanism satisfies (2$\varepsilon$, $\delta$)-joint differential privacy. With probability at least $1-\beta$, it computes an $\eta$-approximate NE, where
$\eta=\zeta+\alpha+E_1+E_2$. Here, the $E_2$ is the approximation brought by \textbf{MW} and
$$E_2  = \tilde {\rm O}{\left( {\frac{{n{\gamma ^2}}}{\varepsilon }\log \left( {\frac{{2kn}}{\beta }} \right)\sqrt {\log \left( k \right)\ln \left( {1/\delta } \right)} } \right)^{1/2}}.$$
\end{thm}
The proof of $\eta$-approximate \emph{ex-post} NE is given in Appendix B.

We now explore the efficiency of our approximate NE.  Intuitively, our approximate parameters $\zeta, E_1,$ and $E_2$ diminish with the increase of number of users with our motivation to study large-scale spectrum sharing. It is apt to set $\alpha$ and $\gamma$ by $\frac{1}{n}$ due to the fact that individual contention probability $q_{i}^d(p_l)$ is not larger than 1. Then,
\begin{equation}
\hspace{-4.5cm}\zeta {\rm{ = }}\sqrt {\frac{{{\rm{8}}\log (2mn)}}{n}},
\end{equation}
\begin{equation}
\vspace{-.2cm}
\hspace{-2.8cm}E_1  = \max _{i\; \to e_1 } \frac{{8(k\log n + \log (4/\beta ))}}{{n\varepsilon }},
\end{equation}
\begin{equation}
\vspace{-.1cm}
\hspace{-0.7cm}E_2  = \widetilde{\rm{O}}\left( {\frac{1}{{n\varepsilon }}\log \left( {\frac{{2kn}}{\beta }} \right)\sqrt {\log \left( k \right)\ln \left( {1/\delta } \right)} } \right)^{1/2}.
\end{equation}

\textbf{Remark 2:}  In future network, the number of spectrum users tends to be very large. However, our approximations $\zeta, E_1$, and $E_2$ have good performance with large user number $n$. For approximation $\zeta$ brought by Lipschitz game and $E_2$ from \textbf{MW}, their scaling law on user number is in $\widetilde{\rm{O}}(\sqrt {\frac{{\log n}}{n}} )$, showing that they become relatively smaller with $n$ growing larger. Similar for the error bound of \textbf{SparCost}, its scaling law on user number is in $\widetilde{\rm{O}}(\frac{{\log n}}{n})$. All these results lead to superior performance in large-scale spectrum sharing. Then, we focus on how number of channels can affect individual utility. The approximation $\zeta, \alpha, E_1,$ and $E_2$ define total error caused by the computation. However, since a user can access only one channel at a time, we measure average approximation of channels.
\begin{equation}
\hspace{-2.3cm}E_1  = \max _{i\; \to e_1 } \frac{{8(k\log n + \log (4/\beta ))}}{{kn\varepsilon }},
\end{equation}
\begin{equation}
E_2  = \widetilde{\rm{O}}\left( {\frac{1}{{kn\varepsilon }}\log \left( {\frac{{2kn}}{\beta }} \right)\sqrt {\log \left( k \right)\ln \left( {1/\delta } \right)} } \right)^{1/2}.
\end{equation}
It is explicit that the scaling law in terms of channel number  $\widetilde{\rm{O}}(\sqrt {\frac{{\log k}{\sqrt{\log k}}}{k}} )$ is smaller with $k$ growing large.

\textbf{Remark 3:}  It should be noted that even if there are not many users who are willing to opt-in, we still reach an approximate NE. Our result of approximate NE does not depend on the number of users who opt-in. Only the efficiency of the approximation can be affected by them. When \textbf{SparCost} chooses a set-valued aggregative function whose corresponding equilibrium is not so close to the expected equilibrium, the approximation suffers.


If we carefully choose the confidence parameter $\beta$ and privacy parameters ($\varepsilon,\delta$), the result can be even better. Hence, we have proved that our approximate NE result performs better in larger-scale spectrum sharing, which is promising for future wireless network.

Finally, we state that our privacy-preserving mechanism is incentive compatible, which means that users are encouraged to opt-in and thus they can get a better payoff.
\begin{thm}
The output of \textbf{MW} is incentive compatible.
\end{thm}
The proof is given in Appendix C.

\subsection{Further Discussion on User Dynamics}

We further discuss the impact of user dynamics on the performance. Users could be mobile or
experiences spectrum mobility \cite{Akyildiz1} in spectrum sharing environments. We here consider the mobility of users in an uncertain environment.
Assume a user has a limited transmission range, his location determines how many channels he can access. We then define spatial $\gamma$-aggregative game $\Gamma  = (\mathcal{N},\mathcal{M},\mathcal{K_D},\mathcal{D},\{ S(\mathcal{N'})_i^d\},\{ U_i(\mathcal{D})\},\gamma,)$. $\mathcal{D}$ is the location set of each user and $\mathcal{K_D}$ is the channels set according to the location set. $\mathcal{N}'$ denotes the conflicting user set.  In our result, users' expected utilities are related to the amount of channels. Therefore, a user's utility can be associated with his location.

As noticed, users could be mobile to
seek for better channel access opportunities. Because the location could affect the utility, users tend to move to places with higher payoff, which are places with more channels or better channel qualities. However, individual throughput declines with the increase of user number. Although we have shown that more users lead to better performance in the large-scale spectrum sharing scenario, the result is in terms of NE and is based on overall utility.  Consequently, these users tend to find a balance between the number of channels and the number of conflicting users.

\begin{figure*}
\centering
\includegraphics[scale=.76]{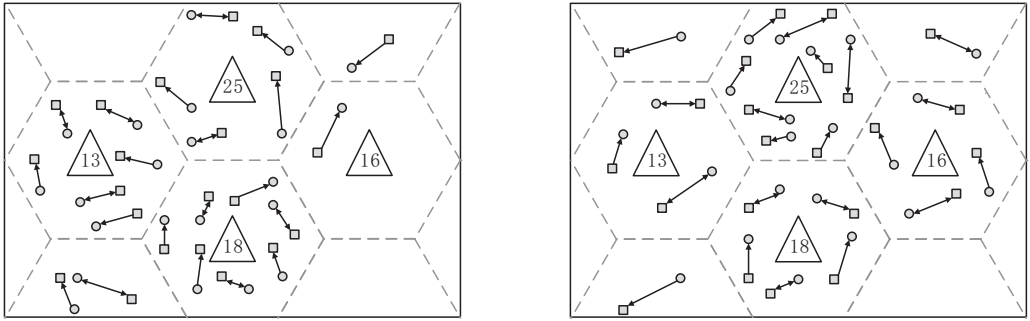}
\caption{The large scale network scenarios}
\label{fig:digraph}
\vspace{-.4cm}
\end{figure*}
Intuitively from our analysis, the private and truthful scaling results in \textbf{Theorem 3, 5, 6, 7, 8} still hold and only have some  constant loss. Simulation results support our conclusion, i.e.,  a user can have better utility when there are more channels in the transmission range and individual usage degrades with the increase  number of competing users. It is also shown that users tend to change their location all the time searching for the balance point. Note that users' changing  their location can be usually modeled by a typical Markov chain \cite{CH}. Its stationary distribution and properties need further investigation and are our future work.

\section{Simulation Results}

We now evaluate the proposed mechanism and its dynamics by simulations. All results are averaged over 1000 runs. We first show our mechanism's performance with different amounts of users and channels, respectively. As for the number of users, we vary it from 500 to 2000 with a step of 100. These conflicting users are randomly deployed in a hexagonal area with side length 500m (Fig. 5). The hexagon represents the base station's signal coverage. Without loss of generality, only one base station exists in the middle of the hexagon and has multiple channels. We set the number of actions $m$ by 50 and the number of channels by 15 on the base station. We also assume the confidence constant $\beta$ to 0.25, the privacy constants $\varepsilon$ to 0.1 and $\delta$ to 0.25. As it did in previous  section, $\gamma$ and $\alpha$ are set by $\frac{1}{n}$. We evaluate the efficiency by showing the trends of $\eta=\zeta+\alpha+E_1+E_2$.

As shown in Fig. 6, the approximation decreases in a speed of $ \frac{\log n}{n}$ with the growing number of users.
Since $\eta$ is the worst case approximation in our mechanism, we show that longer learning periods of $T$ can lead to better performance using the worst case $\eta$ as a benchmark. When we set the learning period $T$ by 10, 20, 30 respectively, Fig. 6 shows that both the starting points and the ends of approximation gradually become smaller with longer periods. Note that the optimal approximation cannot be 0 because of the privacy concern.

We also compare 1000-user hexagon areas whose number of channels varies from 5 to 20, and measure the average effect of channels $\frac{\eta}{k}$. As shown in Fig. 7, the average approximation decreases in a scaling law of $\widetilde{\rm{O}}(\sqrt {\frac{{\log k}{\sqrt{\log k}}}{k}} )$, which indicates that a channel can be better utilized when there are more channels. Fig. 7 also shows the average approximation becomes smaller when learning period of $T$ is larger.

\begin{figure}
\centering
\includegraphics[scale=.55]{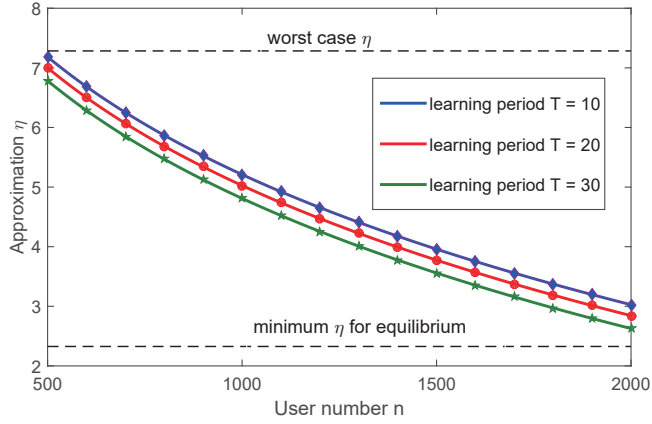}
\caption{Performance on the number of users}
\label{fig:digraph}
\vspace{-.2cm}
\end{figure}

Then, we show the benefit of mediator. When there are 1000 users in one hexagon, we set the ratio of opting-in users to opting-out users by 2:8, 5:5, 8:2, respectively. Consider a Rayleigh fading channel environment, the channel gain $\emph{g}^d_{i}$ is an exponentially distributed random variable. In following simulation, we set user $i$'s spectrum bandwidth $B_{i}^{d}$ by 20MHz and his fixed transmission power $\phi^{d} _{ij}$ by 100mW. The background noise power $\omega _{i}^d$ is -100dBm. Users' contention probabilities are randomly selected from the set \{0.1,0.2,...0.9\} while their mixed strategies are randomly selected from \{0.01,0.02,...,0.99\}. As shown in Fig. 8, the average utility $\mathop \mathbb{E}\nolimits_{\vec P} U_{ij}^d (p_i ,Q_i^d (\vec p,\vec P_d ))$ for opting-out users is less than that $\mathop \mathbb{E}\nolimits_{\overline P} U_{ij}^d (p_i ,Q_i^d (\vec p,\overline P_d ))$ of opting-in users. With more opting-in users, the gap between the two utilities becomes larger, showing that our mechanism is incentive compatible. Noted that the introduction of differential privacy slightly lowers individual utility to prevent users' \emph{types} from being exposed .

\begin{figure}
\centering
\includegraphics[scale=.55]{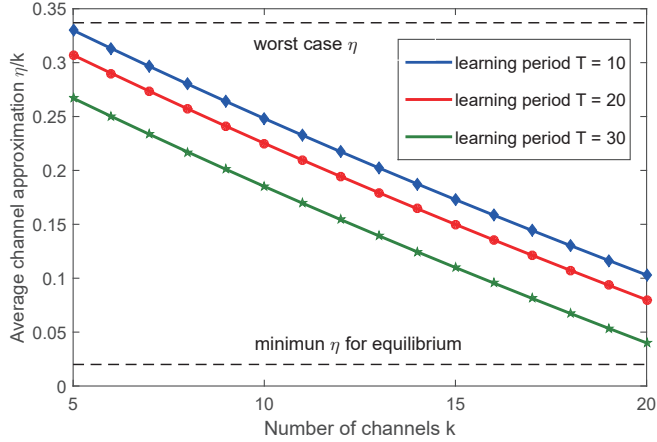}
\caption{Performance on the number of channels}
\label{fig:digraph}
\vspace{-.2cm}
\end{figure}

Finally, we show the dynamics of spectrum users. The left of Fig. 5 is a random distribution of users when there are many base stations. Each base station has its hexagon and users in different hexagons can use the spectrum simultaneously. Such phenomenon is called spatial reuse and is studied in \cite{CH} via a congestion game. The number inside the hexagon denotes the number of channels in the area. We have shown that the average approximation of channels diminishes with the increase of channel number. So users can get a better payoff if the hexagon he is in possesses more available channels. With this in mind, users tend to move to places with more channels. However, individual usage deteriorates with the growth of conflicting users. In Fig. 5, the mobility of spectrum users with learning period $T=20$ is shown. After $T$=100 periods, the left graph turns into the right graph. We
can see from Fig. 5 that some of the users choose to stay in his original
hexagon while others change their locations. Similarly, Fig. 9 shows
the dynamics over time, indicating the impact of the number of channels on the dynamics.  The stationary distribution of whether a user moves or
not needs further investigation.

\begin{figure}
\centering
\includegraphics[scale=.52]{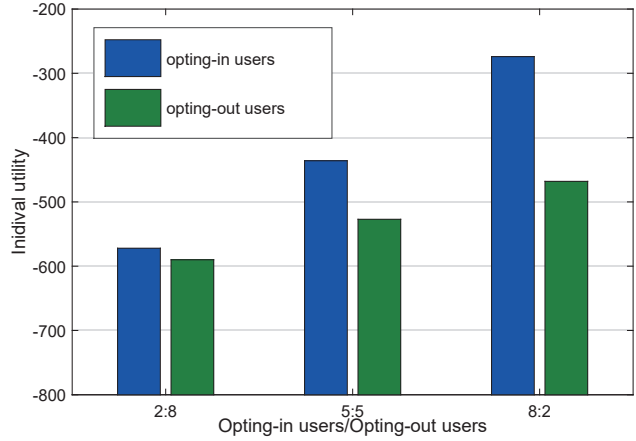}
\caption{Performance of the mediator}
\label{fig:digraph}
\vspace{-.2cm}
\end{figure}

\section{Conclusion and Future Works}
In this paper, we modeled spectrum sharing of large-scale wireless network as a multi-dimensional aggregative game by taking users' heterogeneous impact into account. To the best of our knowledge, this is the first study in large-scale spectrum sharing via aggregative game while concerning privacy and truthfulness. We designed a mediated privacy-preserving and truthful game which admits an   $\eta$-approximate \emph{ex-post} NE. We also showed that the equilibrium has better performance when the spectrum sharing is in larger scale. We demonstrated that the approximation of NE decreases in a speed of $ \frac{\log n}{n}$ with the growing number of users, and  decreases in a scaling law of $\widetilde{\rm{O}}(\sqrt {\frac{{\log k}{\sqrt{\log k}}}{k}} )$ on the channel.  We proved that our game satisfies joint differential privacy and is incentive compatible. Numerical result supports our conclusion. In the future, we plan to investigate the dynamic channel selection game and find the stationary distribution of users' dynamics.

\section*{Appendix A Proof of the accuracy bound for SparCost}

We first prove that $e_1  = \frac{{8\gamma (\log N + \log (4/\beta )}}{\varepsilon }$, then we prove \textbf{SparCost} satisfies $\varepsilon$-differential privacy.
\begin{figure}
\centering
\includegraphics[scale=.47]{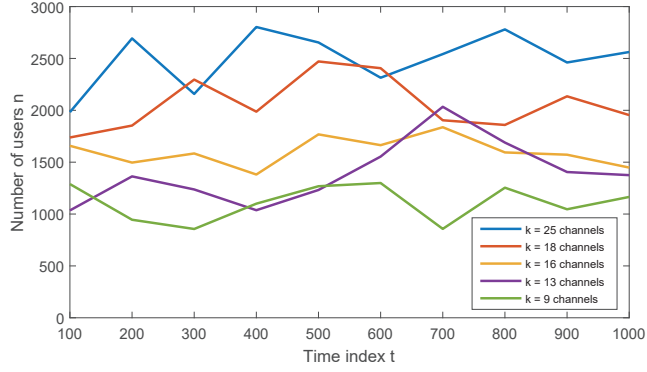}
\caption{User dynamics   with user fluctuation under different number of channels}
\label{fig:digraph}
\vspace{-.2cm}
\end{figure}
For error bound $e_1$, $e_1  = \frac{{8\gamma (\log N + \log (4/\beta )}}{\varepsilon }$.
\begin{IEEEproof}
Let $[N]$ be the set of sequence number $\{1,...,N\}$. Observe that we can prove the theorem by showing that except with probability at most $\frac{\beta}{2}$:
$$ \mathop {\max}\nolimits_{n\in[N]} |v_n| + |T - \mathord{\buildrel{\lower3pt\hbox{$\scriptscriptstyle\frown$}}\over T}| \leq e_1.$$
If this is the case, then for any output $\theta_n$,
 $$ a_n(\widehat Q_{ij}^d( \vec p,\vec P_d))+v_n \le \mathord{\buildrel{\lower3pt\hbox{$\scriptscriptstyle\frown$}}\over T} \le T+|T-\mathord{\buildrel{\lower3pt\hbox{$\scriptscriptstyle\frown$}}\over T}|, $$
or in other words:
$$ a_n(\widehat Q_{ij}^d( \vec p,\vec P_d)) \le T+|T-\mathord{\buildrel{\lower3pt\hbox{$\scriptscriptstyle\frown$}}\over T}|+|v_n| \le T+e_1. $$
Similarly, for any output $\bot$,
 $$ a_n(\widehat Q_{ij}^d( \vec p,\vec P_d)) > \mathord{\buildrel{\lower3pt\hbox{$\scriptscriptstyle\frown$}}\over T} \ge T-|T-\mathord{\buildrel{\lower3pt\hbox{$\scriptscriptstyle\frown$}}\over T}|-|v_n| \ge T-e_1. $$
For any $n< N$ : $a_n(\widehat Q_{ij}^d( \vec p,\vec P_d)) > T-e_1 > T+|v_n|+|T-\mathord{\buildrel{\lower3pt\hbox{$\scriptscriptstyle\frown$}}\over T}|$, and also : $a_n(\widehat Q_{ij}^d( \vec p,\vec P_d))+v_n \ge \mathord{\buildrel{\lower3pt\hbox{$\scriptscriptstyle\frown$}}\over T}$, meaning the output is $\bot$. Therefore the algorithm does not halt before all $N$ candidates are checked.

 Recall that if $X$ satisfies Laplace distribution: $X \sim Lap(b)$, then: $Pr[|X| \ge t \cdot b] = exp(-t)$. Therefore,
$$ Pr[|T-\mathord{\buildrel{\lower3pt\hbox{$\scriptscriptstyle\frown$}}\over T}| \ge \frac{e_1}{2}] = exp(- \frac{\epsilon e_1}{4\gamma}). $$
We set this quantity by at most $\beta/4$, we then find $e_1 \ge \frac{4 \log{(4/\beta)}}{\epsilon}$.
 Similarly, by a union bound, we have:
$$ Pr[ \mathop{\max}\nolimits_{i \in |N|}|v_n| \ge \frac{e_1}{2} ] \le N \cdot exp(- \frac{\epsilon e_1}{8\gamma}). $$
We set this quantity by at most $\beta/4$ as well, we then find $ e_1 \ge \frac{8\gamma(\log{(4/\beta)}+\log{N})}{\epsilon}. $ These two claims combine to prove $e_1  = \frac{{8\gamma (\log N + \log (4/\beta )}}{\varepsilon }.$
\vspace{-.1cm}
\end{IEEEproof}

Now, we turn to prove that \textbf{SparCost} satisfies $\varepsilon$-differential privacy.
\begin{IEEEproof}
At first, we need to redefine some of the notations. Let $a_n(Q, \widehat Q)=a_n(\widehat Q_{ij}^d( \vec p,\vec P_d))$. $Q$ is the aggregator from submitted mixed strategy while $\widehat Q$ is still set-valued aggregative function. Let $Q'$ be the aggregator when one of the users submits another mixed strategy. Let $\theta$ and $\theta'$ be the output of \textbf{SparCost} when the input is $Q$ and $Q'$ respectively. Noted that $\theta$ here includes both desired cost and $\bot$.

In following analysis, we fixed values of Laplace noise $v_1$,...,$v_{N-1}$. For $v_N$ and noisy threshold $\mathord{\buildrel{\lower3pt\hbox{$\scriptscriptstyle\frown$}}\over T}$, we take probabilities over the randomness of them. Define $g(Q)$ as the minimum noisy value of $a_1(Q, \widehat Q)$,..., $a_{N-1}(Q, \widehat Q)$ and so:
$$\vspace{-0.1cm}g(Q)=\min _{n < N} (a_n(Q, \widehat Q)-v_n).\vspace{-.1cm}$$
We also use $\mathop{\Pr}[\mathord{\buildrel{\lower3pt\hbox{$\scriptscriptstyle\frown$}}\over T}=t]$ as the probability distribution function of $\mathord{\buildrel{\lower3pt\hbox{$\scriptscriptstyle\frown$}}\over T}, \mathop {\Pr }[v_N=v]$ for the noise and $\mathop{\Pr}[a_n(Q, \widehat Q)=a]$ for the output. Let $I[x]$ be the indicator of event $x$. We then have:
$$\hspace{-0.cm} \mathop {\Pr }\limits_{\mathord{\buildrel{\lower3pt\hbox{$\scriptscriptstyle\frown$}}\over T},v_N } [\theta = a] = \mathop {\Pr }\limits_{\mathord{\buildrel{\lower3pt\hbox{$\scriptscriptstyle\frown$}}\over T},v_N }[\mathord{\buildrel{\lower3pt\hbox{$\scriptscriptstyle\frown$}}\over T}\le g(Q) \quad and \quad a_N(Q, \widehat Q)-v_N \le \mathord{\buildrel{\lower3pt\hbox{$\scriptscriptstyle\frown$}}\over T} ]
=\mathop {\Pr }\limits_{\mathord{\buildrel{\lower3pt\hbox{$\scriptscriptstyle\frown$}}\over T},v_N }[\mathord{\buildrel{\lower3pt\hbox{$\scriptscriptstyle\frown$}}\over T}\in (a_N(Q, \widehat Q)-v_N,g(Q))] $$
$$\hspace{-0.25cm}= \int_{ - \infty }^{ + \infty } {\int_{ - \infty }^{ + \infty } {\mathop {\Pr }[v_N=v]} } {\mathop {\Pr }[\mathord{\buildrel{\lower3pt\hbox{$\scriptscriptstyle\frown$}}\over T}=t]}
*I[t\in (a_N(Q, \widehat Q)-v,g(Q))]dvdt$$
$$\hspace{-10.cm}= \omega.
\vspace{-.3cm}$$

We now change some of the variables:
$$\mathord{\buildrel{\lower3pt\hbox{$\scriptscriptstyle\frown$}}\over v}=v-g(Q)+g(Q')-a_N(Q', \widehat Q)+a_N(Q, \widehat Q), \mathord{\buildrel{\lower3pt\hbox{$\scriptscriptstyle\frown$}}\over t}=t-g(Q)+g(Q').
\vspace{-.2cm}$$
Then we have:
$$\omega= \int_{ - \infty }^{ + \infty } {\int_{ - \infty }^{ + \infty } {\mathop {\Pr }[v_N=\mathord{\buildrel{\lower3pt\hbox{$\scriptscriptstyle\frown$}}\over v}]} } {\mathop {\Pr }[\mathord{\buildrel{\lower3pt\hbox{$\scriptscriptstyle\frown$}}\over T}=\mathord{\buildrel{\lower3pt\hbox{$\scriptscriptstyle\frown$}}\over t}]}*I[t-g(Q)+g(Q')\!\!\in\!\! (a_N(Q',\! \widehat Q)-v-g(Q)+g(Q')\!,\!g(Q))]dvdt$$
$$\hspace{-4.0cm}=\int_{ - \infty }^{ + \infty } {\int_{ - \infty }^{ + \infty } {\mathop {\Pr }[v_N=\mathord{\buildrel{\lower3pt\hbox{$\scriptscriptstyle\frown$}}\over v}]} } \mathop {\Pr }[\mathord{\buildrel{\lower3pt\hbox{$\scriptscriptstyle\frown$}}\over T}=\mathord{\buildrel{\lower3pt\hbox{$\scriptscriptstyle\frown$}}\over t}]*I[t \in (a_N(Q', \widehat Q)-v,g(Q'))]dvdt$$
$$\hspace{-1.7cm}\le \int_{ - \infty }^{ + \infty } {\int_{ - \infty }^{ + \infty }\exp(\frac{\varepsilon}{2}) {\mathop {\Pr }[v_N=\mathord{\buildrel{\lower3pt\hbox{$\scriptscriptstyle\frown$}}\over v}]} }\exp(\frac{\varepsilon}{2}) \mathop {\Pr }[\mathord{\buildrel{\lower3pt\hbox{$\scriptscriptstyle\frown$}}\over T}=\mathord{\buildrel{\lower3pt\hbox{$\scriptscriptstyle\frown$}}\over t}]*I[t \in (a_N(Q', \widehat Q)-v,g(Q'))]dvdt$$
\vspace{-.7cm}
$$\hspace{-7.3cm}=\exp(\varepsilon)\mathop {\Pr }\limits_{\mathord{\buildrel{\lower3pt\hbox{$\scriptscriptstyle\frown$}}\over T},v_N }[\mathord{\buildrel{\lower3pt\hbox{$\scriptscriptstyle\frown$}}\over T} <g(Q') and a_N(Q', \widehat Q)-v_N \le \mathord{\buildrel{\lower3pt\hbox{$\scriptscriptstyle\frown$}}\over T}]$$
\vspace{-.7cm}
$$\hspace{-11.5cm}=\exp(\varepsilon)\mathop {\Pr }\limits_{\mathord{\buildrel{\lower3pt\hbox{$\scriptscriptstyle\frown$}}\over T},v_N }[\theta'=a].$$
Thus completes our proof that \textbf{SparCost} satisfies $\varepsilon$-differential privacy.
\vspace{-.4cm}

\end{IEEEproof}

\section*{Appendix B Proof of $\eta$-approximate NE}

We now prove that our game is $\eta$-approximate, where $\eta=\zeta+\alpha+E_1+E_2$.

In the \textbf{MW}, the channel connection probabilities on the billboard is proceeded by both \textbf{SparCost} and \textbf{REXP}. According to composition theorem \cite{DR}, these probabilities possess (2$\varepsilon$, $\delta$)-differential privacy. With regard to Billboard lemma, our result satisfies (2$\varepsilon$, $\delta$)-joint differential privacy.

According to Theorem 1 of Lipschitz Game, the assumption that our game has $n$ players, $m$ actions, and satisfies Lipschitz constant of $\gamma$ ensures that our game converges to $\gamma \sqrt {{\rm{8}}n\log (2mn)}$ -approximate NE. This NE does not concern any accuracy issue from privacy and algorithm.

Now we consider the inaccuracy induced from privacy and algorithm. From Theorem 3, we have
$$E_1= \max _{i\ \to e_1 } \frac{{8\gamma (k\log \frac{{n\gamma }}{\alpha } + \log (4/\beta ))}}{\varepsilon }.$$
\vspace{-.2cm}
 We then explore the accuracy of \textbf{MW} which is
\begin{equation}
E_2  = \tilde {\rm O}{\left( {\frac{{n{\gamma ^2}}}{\varepsilon }\log \left( {\frac{{2kn}}{\beta }} \right)\sqrt {\log \left( m \right)\ln \left( {1/\delta } \right)} } \right)^{1/2}}.
\end{equation}

\begin{IEEEproof}
The \textbf{MW} algorithm gives a no-regret guarantee for each user $i$:
$$  \frac{1}{T}\sum\nolimits_t {\left\langle {P_i^t,q_{i}(p_i)} \right\rangle}\le \mathop {\min }\nolimits_{{P_i} } \frac{1}{T}\sum\nolimits_t {\left\langle {P_i^t,q_{i}(p_i)} \right\rangle }  + \eta  + \frac{{\log (k)}}{{{T_\eta }}}= \mathop {\min }\nolimits_{{P_i}} \frac{1}{T}\sum\nolimits_t {\left\langle {P_i^t,q_{i}(p_i)} \right\rangle }  + \frac{E_2 }{{2n\gamma }},$$
where ${\left\langle {P_i^t,q_{i}(p_i)} \right\rangle }$ is the input and output of \textbf{REXP}. We study the effect of $q_{i}^d(p_i)$ for all channels and omit $d$. $\lambda^t$ in following proof is from the score function in \textbf{REXP}. Let $q(p)$ be the set for channel contention probability
of all users on all channels, i.e., $q(p)=\{\overline{q}_{1}^d(p_1),...,\overline{q}_{n}^d(p_n)\}$. Thus, the joint play of all $n$ agents satisfies
\begin{equation}
\vspace{-.1cm}
\frac{1}{T}\sum\nolimits_t {\left( {\gamma \left\langle {P^t,q(p)} \right\rangle  - {\lambda^t}} \right)}\le \mathop {\min }\nolimits_{P} \frac{1}{T} \sum\nolimits_t {\left( {\gamma \left\langle {P^t,q(p)} \right\rangle   - {\lambda^t}} \right)}  + E_2 /2.
\end{equation}
It is easy to see that
\begin{equation}
\vspace{-.2cm}
\mathop {\min }\nolimits_{P} \frac{1}{T}\sum\nolimits_t {\left( {\gamma \left\langle \vec {P^t,q(p)} \right\rangle  - {\lambda^t}} \right)}  \le 0.
\end{equation}
So from above we know,
\begin{equation}
\label{proof1}
\vspace{-.1cm}
\frac{1}{T}\sum\nolimits_t {\left( {\gamma \left\langle {P^t,q(p)} \right\rangle  -{\lambda^t}} \right)}  \le E_2 /2.
\end{equation}

By Theorem 4, with probability at least $1-\frac{\beta}{2}$, exponential mechanism gives
$$\hspace{-.cm}\sum\nolimits_t {\left( {\gamma \left\langle {P^t,q(p)} \right\rangle  - {\lambda^t}} \right)} \ge \mathop {\max }\nolimits_{\left( {q,\lambda} \right)} \sum\nolimits_t {\left( {\left( {\gamma \left\langle {P^t,q(p)} \right\rangle  - {\lambda_d^t}} \right) - \frac{{2\gamma \log \left( {\frac{{2kT}}{\beta }} \right)}}{{{\varepsilon _0}}}} \right)}.
\vspace{-.2cm}$$
And so,
\begin{equation}
\label{proof2}
\vspace{-.1cm}
\frac{1}{T}\sum\nolimits_t {\left( {\gamma \left\langle {P^t,q(p)} \right\rangle  - {\lambda^t}} \right)}\ge \mathop {\max }\nolimits_{\left( {q,\lambda} \right)} \frac{1}{T}\left[ {\sum\nolimits_t {\left( {\gamma \left\langle {P^t,q(p)} \right\rangle  - {\lambda^t}} \right)} } \right] - \frac{{2\gamma \log \left( {\frac{{2kT}}{\beta }} \right)}}{{{\varepsilon _0}}}.
\end{equation}

Combining above (\ref{proof1}) and (\ref{proof2}) with the definition of $\overline {P}$, we get
$$\mathop {\max }\nolimits_{(q,\lambda)} \left( {\gamma \left\langle {\overline {P},q(p)} \right\rangle  - \lambda} \right) = \mathop {\max }\nolimits_{\left( {q,\lambda} \right)}
\frac{1}{T}\sum\nolimits_t {\left( {\gamma \left\langle {P^t,q(p)} \right\rangle \!- \! {\lambda^t}}\right)}\le \frac{{2\gamma \log \left( {\frac{{2kT}}{\beta }} \right)}}{{{\varepsilon _0}}}  + E_2 /2 \le E_2 ,$$
as long as $\frac{2\gamma \log \left(\frac{2kT}{\beta} \right)}{\varepsilon _0} \leq E_2 /2$. Plugging in parameter $\varepsilon _0$ from differential privacy, we then have
\begin{equation}
\vspace{-.1cm}
{E_2 ^2} \ge \frac{{32\sqrt 2 n{\gamma ^2}\log \left( {\frac{{2kT}}{\beta }} \right)\sqrt {\log k\ln \left( {\frac{1}{\delta }} \right)} }}{\varepsilon }.
\end{equation}

Plugging in $T$, we get our desired bound
\begin{equation}
\vspace{-.4cm}
E_2  = \tilde {\rm O}{\left( {\frac{{n{\gamma ^2}}}{\varepsilon }\log \left( {\frac{{2kn}}{\beta }} \right)\sqrt {\log \left( m \right)\ln \left( {1/\delta } \right)} } \right)^{1/2}}.
\end{equation}
\end{IEEEproof}

Combining above, we have completed the proof. Now we have $\eta=\zeta+\alpha+E_1+E_2$.

\section*{Appendix C Proof of Incentive Compatible}
The large-scale spectrum sharing game we are discussing here contains two undesirable deviations. One is opting out of the mediated game. The other one is lying or not faithfully following the suggestion. However, it is incentive compatible. In fact, our mechanism encourages users to opt-in and to behave themselves, and the corresponding proof is as follows.

\begin{IEEEproof}
We first prove why users who tend to lie or not to truthfully follow the solution should  play faithfully. Let $\overline P$ be the mixed strategy set for all users under NE. $P_i$ and $\overline P_{-i}$ are corresponding mixed strategies for user $i$ and users other than $i$, respectively. Let $x_i$ be user $i$'s type. We concern the type in the utility in the proof and assume that type $x_i$ needs the corresponding $P_i$ to get the desired utility. Since our game converges to $\eta$-approximate NE, we have,
\begin{equation}
\label{APP1}
\vspace{-.2cm}
\mathop \mathbb{E}\nolimits_{x_i ,x_{ - i} }U_{ij} (p_i ,Q_i^d (\vec p,\overline P)) \ge\mathop \mathbb{E}\nolimits_{x_i ,x_{ - i} } U_{ij} (p_i' ,Q_i^d ((p_i',\vec p_{-i}),(P_i',\overline P_{-i})) - \eta,
\end{equation}
where $p'_i$ denotes the action with the best utility and $ p_i$ is an action of aggregative best response. $P_i'$ and $P_i$ are corresponding mixed strategies. In other words, action $p'_i$ leads to exact NE while action $ p_i$ leads to a NE in terms of aggregative best response. The latter is no worse than the former by at most $\eta$. (\ref{APP1}) provides the reason why users are willing to opt-in. Only in this way can they get a better payoff via reaching a NE.


Then, we invoke the privacy condition to explain why users are willing to opt-in telling the true mixed strategies and follow the recommendation. Our result satisfies (2$\varepsilon$, $\delta$)-joint differential privacy, so
$$\mathop \mathbb{E}\nolimits_{x_i ,x_{ - i} }U_{ij} (p_i ,Q_i^d (\vec p,\overline P))$$
$$\hspace{-2cm}\ge\mathop \mathbb{E}\nolimits_{x_i ,x_{ - i} } U_{ij} (p_i' ,Q_i^d ((p_i',\vec p_{-i}),(P_i',\overline P_{-i})) - \eta$$
$$  \ge \exp ( - 2\varepsilon )\mathop \mathbb{E}\nolimits_{x_i ',x_{ - i} } U_{ij} (p_i' ,Q_i^d ((p_i',\vec p_{-i}),(P_i',\overline P_{-i})) - \eta  - \delta$$
$$\hspace{-0.5cm}\ge \mathop \mathbb{E}\nolimits_{x_i ',x_{ - i} } U_{ij} (p_i' ,Q_i^d ((p_i',\vec p_{-i}),(P_i',\overline P_{-i})) - \eta  - \delta  - 4L\varepsilon,\vspace{-0.2cm}$$
where $L$ is the upper bound of the utility function. Noted that $x_i '$ means user $i$'s type is not mapped into aggregative best-response to  $x_{-i} $. This property comes from joint differential privacy rather than the standard. The last inequality is the result of inequality $e^\varepsilon   \le 1 + 2\varepsilon.$
 Then, consider the analysis of $\beta$-probability event that the mechanism fails to output an approximate NE, we have:
\begin{equation}
\mathop \mathbb{E}\nolimits_{x_i ,x_{ - i} }U_{ij} (p_i ,Q_i^d (\vec p,\overline P)) \ge   \mathop \mathbb{E}\nolimits_{x_i ',x_{ - i} } U_{ij} (p_i' ,Q_i^d ((p_i',\vec p)_{-i}),(P_i',\overline P_{-i})) - \eta  - \delta  - 4L\varepsilon  - L\beta.
\end{equation}
This completes the proof.
\end{IEEEproof}



\begin{thebibliography}{1}


\bibitem{MY}
M. Y.  Liu and Y.  Wu, ``Spectum sharing as congestion games," in \emph{IEEE Communication, Control, and Computing, 46th Annual Allerton Conference}, pp. 1146-1153, 2008.
\bibitem{CH}
X. Chen and J. Huang, ``Distributed spectrum access with spatial reuse," \emph{IEEE J. Sel. Areas Coommun.}, vol. 31, no. 3, pp. 593-603. 2013.

\bibitem{NZD}
A. Nasipuri, J. Zhuang  and S.R. Das, ``A multichannel CSMA MAC protocol for multihop wireless networks," in \emph{IEEE Wireless Communications and Networking Conference (WCNC)}, pp. 1402-1406, 1999.
\bibitem{NH1}
D. Niyato and E. Hossain, ``A noncooperative game-theoretic framework for radio resource management in 4G heterogeneous wireless access networks," \emph{IEEE Trans. Mobile Computing}, vol. 7, no. 3, pp. 332-345, 2008.

\bibitem{JENSEN}
M. K. Jensen, ``Aggregative games and best-reply potentials," \emph{Economic theory}, vol. 43, no. 1, pp. 45-66, 2010.
\bibitem{NASH}
J. Nash, ``Non-cooperative games," \emph{Annals of mathematics}, pp. 286-295, 1951.
\bibitem{RR}
R. M. Rogers, A. Roth, ``Asymptotically truthful equilibrium selection in large congestion games," in \emph{Proc. ACM conference on Economics and computation}, pp. 771-782, 2014.
\bibitem{EPT}
R. Etkin and  A. Parekh, et al., ``Spectrum sharing for unlicensed bands," \emph{IEEE J. Sel. Areas Commun.}, vol. 25, no. 3, pp. 517-528, 2007.
\bibitem{HUANGQ}
Q. Huang, Y. Tao and F. Wu, ``Spring: A strategy-proof and privacy preserving spectrum auction mechanism," in \emph{The IEEE International Conference on Computer Communications (Infocom)}, pp. 827-835, 2013.
\bibitem{Naor99}
M. Naor, B. Pinkas, R. Sumner, ``Privacy preserving auctions and mechanism design," in \emph{Proc. ACM conference on Electronic commerce}, pp. 129-139, 1999.
\bibitem{DR}
C. Dwork and A. Roth, ``The algorithmic foundations of differential privacy," \emph{Foundations and Trends in Theoretical Computer Science}, vol. 9, no. 3-4, pp. 211-407, 2014.
\bibitem{ZLWS}
R. Zhu, Z. Li, F. Wu, et al., ``Differentially private spectrum auction with approximate revenue maximization," in \emph{Proc. ACM international symposium on mobile Ad Hoc networking and Computing (MobiHoc)}, pp. 185-194, 2014.
\bibitem{ARORA}
S. Arora, E. Hazan, S. Kale, ``The Multiplicative Weights Update Method: a Meta-Algorithm and Applications," \emph{Theory of Computing}, vol.8, no.1, pp. 121-164, 2012.
\bibitem{NH2}
D. Niyato, E. Hossain, ``Competitive spectrum sharing in cognitive radio networks: a dynamic game approach," \emph{IEEE Trans. Wireless Commun.}, vol. 7, no. 7, pp. 2651-2660, 2008.
\bibitem{HUANG}
X. Chen, J. Huang, ``Spatial spectrum access game: Nash equilibria and distributed learning," in \emph{Proc. ACM international symposium on Mobile Ad Hoc Networking and Computing(MobiHoc)}, pp. 205-214, 2012.
\bibitem{NH}
D. Niyato, E. Hossain, ``Dynamics of network selection in heterogeneous wireless networks: an evolutionary game approach," \emph{IEEE Trans. Vehicular Technology}, vol. 58, no. 4, pp. 2008-2017, 2009.
\bibitem{Huang06}	
J. Huang, R. A. Berry, and M. L.  Honig, ``Auction-based spectrum sharing. Mobile Networks and Applications," \emph{Mobile Networks and Applications}, vol. 11, no. 3, pp. 405-418, 2006.
\bibitem{ROSE}
R. W. Rosenthal, ``A class of games possessing pure-strategy Nash equilibria," \emph{International Journal of Game Theory}, vol. 2, no. 1, pp. 65-67, 1973.
\bibitem{AGT}
T. Roughgarden, E.  Tardos, et al., \emph{Algorithmic game theory}. Vol. 1. Cambridge: Cambridge University Press, 2007.
\bibitem{AMD}
N. Nisan and A. Ronen, ``Algorithmic mechanism design," in \emph{Proc. ACM symposium on Theory of computing (STOC)}, pp. 129-140, 1999.
\bibitem{CLW}
G. Cui, M. Li, et al., ``Analysis and evaluation framework based on spatial evolutionary game theory for incentive mechanism in peer-to-peer network," in \emph{IEEE Trust, Security and Privacy in Computing and Communications Conference (TrustCom)}, pp. 287-294, 2012.
\bibitem{FCZ}
X. Feng, Y. Chen, et al., ``TAHES: A truthful double auction mechanism for heterogeneous spectrums," \emph{IEEE Trans. Wireless Commun.}, vol. 11, no. 11, pp. 4038-4047, 2012.
\bibitem{ZS}
R. Zhu and  K. G. Shin, ``Differentially private and strategy-proof spectrum auction with approximate revenue maximization," in \emph{The IEEE International Conference on Computer Communications (Infocom)}, pp. 918-926, 2015.
\bibitem{SMKT}
F. McSherry, K. Talwar, ``Mechanism design via differential privacy," in \emph{IEEE Symposium on Foundations of Computer Science (FOCS)}, pp. 94-103, 2007.
\bibitem{DMNA}	
D. Cynthia, F. McSherry, et al., ``Calibrating noise to sensitivity
in private data analysis," in \emph{Theory of Cryptography Conference}, Springer, pp. 265-284, 2006.
\bibitem{KPRU}
M. Kearns, M.  Pai, et al., ``Mechanism design in large games: Incentives and privacy," in \emph{Proc. ACM conference on Innovations in theoretical computer science}, pp. 403-410, 2014.
\bibitem{Babi}
Y. Babichenko,  ``Best-reply dynamic in large aggregative games," SSRN abstract 2210080, 2013.
\bibitem{AS}
Y. Azrieli and  E. Shmaya, ``Lipschitz games," \emph{Mathematics of Operations Research}, pp. 350-357, 2013.
\bibitem{DRV}
C. Dwork, G. N. Rothblum, et al., ``Boosting and differential privacy," in \emph{IEEE Symposium on Foundations of Computer Science (FOCS)}, pp. 51-60, 2010.
\bibitem{MS}
R. B. Myerson and M. A. Satterthwaite, ``Efficient mechanisms for bilateral trading," \emph{Journal of economic theory}, vol. 29, no. 2, pp. 265-281, 1983.
\bibitem{HHRRW}
J. Hsu,  Z.  Huang, et al., ``Private matchings and allocations," in \emph{ Proc. ACM Symposium on Theory of Computing (STOC)}, pp. 21-30, 2014.

\bibitem{Akyildiz1}
I. F. Akyildiz, W. Y. Lee  and K. R. Chowdhury, ``CRAHNs: Cognitive radio ad hoc networks," \emph{ad hoc networks journal}, vol.7, no.5, pp. 810-836, 2009.
\end{thebibliography}
\end{document}